\RequirePackage{fix-cm}
\documentclass[smallextended]{svjour3}       
\smartqed  
\usepackage{amsmath}
\usepackage{geometry}
\journalname{Foundations of Physics}
\renewcommand{\eqref}[1]{Eq.\ \ref{#1}}

\begin{document}

\title{What is quantum mechanics? A minimal formulation}

\titlerunning{What is quantum mechanics?}        

\author{R. Friedberg         \and
       P. C. Hohenberg 
}

\institute{R. Friedberg \at
             Department of Physics, Columbia University, New York, NY USA \\
             \email{rfriedberg1@nyc.rr.com}             \\
          P.C. Hohenberg \at
            Department of Physics,New York University, New York, NY USA \\
            \email{pierre.hohenberg@nyu.edu}
}

\date{Received: \today}

\maketitle

\begin{abstract}
This paper presents a minimal formulation of nonrelativistic quantum mechanics, by which is meant a formulation which describes the theory in a succinct, self-contained, clear, unambiguous and of course correct manner. The bulk of the presentation is the so-called \lq microscopic theory' (MIQM), applicable to any closed system $S$ of arbitrary size $N$, using concepts referring to $S$ alone, without resort to external apparatus or external agents. An example of a similar minimal microscopic theory is the standard formulation of classical mechanics, which serves as the template for a minimal quantum theory. The only substantive assumption required is the replacement of the classical Euclidean phase space by Hilbert space in the quantum case, with the attendant all-important phenomenon of quantum incompatibility. Two fundamental theorems of Hilbert space, the Kochen-Specker-Bell theorem and Gleason's theorem, then lead inevitably to the well-known Born probability rule. For both classical and quantum mechanics, questions of physical implementation and experimental verification of the predictions of the theories are the domain of the macroscopic theory, which is argued to be a special case or application of the more general microscopic theory.

\end{abstract}

\section{\label{1}	Introduction}

The numerous discussions of quantum foundations that exist in the literature, beginning with the first formulations of the theory over 90 years ago, can be thought of as addressing the following questions:

 A.	What is quantum mechanics? (hereafter QM), i.e. how should one \emph{formulate} the theory in a clear and logically complete way?

B.	Is QM the \lq \emph{whole truth'} with respect to experimental consequences?

C.	What are the physical consequences of the formulation in A? This last question should, strictly speaking, not be classified under quantum foundations, since it is simply what physics research is about. We nevertheless include it in our list, since many discussions of quantum foundations do not distinguish issues belonging to this category from the other questions on our list.

D. How should one interpret QM? Under this large rubric we classify: justifying the assumptions that enter into the formulation in A; examining alternate assumptions and formulations; exploring the implications of QM for philosophy, cognitive science and psychology.

The present paper is exclusively concerned with answering question A. The other questions can be divided into two groups, those that are the stuff of physics research, namely B and C, and everything else, which we have denoted as \lq interpretation'. These issues, classified under D, may be of considerable general interest, but in our view they are unnecessary for the purpose of answering question A, \lq What is QM?'  As mentioned above, most discussions of quantum foundations in the literature fail to distinguish between the different classes of questions we have listed, to the detriment of a clear understanding of the physics.

Note that we do not claim great originality for our minimal formulation of QM. There are so many different formulations and interpretations in the nearly century-long history of the subject that it would be surprising if ours were fundamentally new or if all the others were wrong in any real sense. What we have attempted here is to create a guide and summary of those portions of this huge literature that constitute a minimal and unambiguous answer to the question \lq What is QM?'

The present minimal formulation of quantum mechanics supersedes our earlier work presented in \cite{fh1}, hereafter FH1. It strives to discard \lq excess baggage' present in the earlier version and in many other formulations of QM as well. Although the present work is in some sense a revision of the earlier paper, the text is self-contained and should not require reading the much longer and more elaborate earlier paper.

Section \ref{2} is a brief discussion of classical mechanics (hereafter CM), formulated in such a way as to be easily adapted to the quantum case. Section \ref{3} constitutes the heart of our minimal formulation of QM, the so-called \lq microscopic theory' (MIQM), which characterizes a closed system $S$ of size $N$, purely in terms of quantities defined in $S$. Essentially all the features of QM are consequences of the crucial assumption that the phase space of CM is replaced by a Hilbert space in QM, but the basic formulation is otherwise the same in the two cases. Section \ref{4} addresses the connection of the theory to experiment, i.e. the implementation of the theory in terms of macroscopic preparation and measurement apparatus. This is what we call the macroscopic theory (MAQM), which is a special case of the more general microscopic theory. Section \ref{5} is a brief discussion of some of the most popular formulations and interpretations of QM, seen from the point of view of our minimal formulation. It is argued that such formulations should either be considered as phenomenologies, or as the underlying fundamental theory leading to those phenomenologies. In the latter case, our claim is that most fundamental earlier theories contain considerable material that, however interesting, does not need to be included in a minimal formulation. Section {6} concludes with a summary and with the expressed hope that the Foundations of QM can be laid to rest alongside the Foundations of CM.

\section{\label{2}	Classical Mechanics}
\vskip 10pt
The example of CM provides an illustration of what we mean by a minimal formulation, as well as a template for our own formulation of QM. Question A above, when referring to CM, consists of two parts. The most important one is what we call the microscopic theory (MICM), which is just the textbook description of the theory, applicable to any closed system with an arbitrary number $N$ of degrees of freedom.  As explained below, MICM may be supplemented by a second part, a macroscopic theory (MACM), but this is usually implicit rather than explicit in standard formulations of CM.
\vskip 10pt

\subsection{Microscopic theory (MICM)}\label{2.1}

The basic microscopic theory itself consists of two parts: an ontology and predictions.
\vskip 10pt
\underline{Ontology}

In its simplest form, the ontology is expressed entirely in terms of a Euclidean phase space of $6N$ dimensions, for a closed system with $N$ particles, each one of which has position $q$ and momentum $p$. The ontology consists of system points in phase space, which form the \emph{states} of the system, and (Borel) subsets of phase space, which form the \emph{properties} of the system. Pure states are represented by single points $x$, and mixed states are represented by sets of points $\{x_i\}$ with a probability function $\rho(x_i)$. The time dependence of states is determined by equations of motion governed by the Hamiltonian $H$.

\vskip 10pt
\underline{Predictions}

The predictions of the theory can be expressed entirely in information theoretic terms, by introducing a truth function $\mathcal{T}_x(A)$ acting on properties (denoted by $A$), which can be either true ($\mathcal{T}_x(A) = \textrm{T}$), or false ($\mathcal{T}_x(A) = \textrm{F}$). For a pure state, the property $A$ is true at time $t$ if the representative point $x(t)$ of the state is in the subset of phase space represented by $A$, and it is false otherwise. It follows that the truth and falsehood of properties, which are the essential physical predictions of the theory, depend entirely on the existence (and location) of the state. We can say they are \emph{contextual} to the state. Until a state has been designated (or prepared), all properties are \emph{undetermined}.

For a mixed state, the probability of $A$ is found by summing the probabilities (or integrating the probability density) of all representative points that fall within the subset of phase space represented by $A$. In this probabilistic case it is statements about $\rho$ that are the predictions of the theory. In Appendix A, Kolmogorov's set-theoretic formulation of probability theory is presented and shown to be closely related to classical logic. It follows that the probability function $\rho(x)$ can be thought of as a \emph{distributed truth function}, obtained by combining strict truth functions ${\mathcal T}_x$ according to the weights given by $\rho(x)$.

For pure states, we make the following observation: we could regard the truth functions as primary and the points in phase space as derived.  If we define a pure state to be a truth function $\mathcal{T}_x$ on properties, we must require that the values of $\mathcal{T}_x(A)$ for different $A$ are related by the rules of Aristotelian logic, with $A$ OR $B$ represented by the union $A \cup B$  and NOT $A$ by the complement of $A$.  We could then obtain the phase point corresponding to the state by taking the intersection of all true properties, i.e. properties  $A$ having $\mathcal{T}_x(A) = \textrm{T}$ (see Appendix A). This will be useful when we develop the analogy between CM and QM. For classical mixed states we can also retrieve the probability density $\rho(x)$ from the probability density of properties $\mathcal {M}(A)$,  as the greatest lower bound $\sqcap$ of probabilities of properties possessed by $x$. There is thus a one-to-one correspondence between pure or mixed classical states, defined as functions of points in phase space, and truth or probability functions defined over properties (subsets of phase space points).

\vskip 10pt
\noindent
\underline{Microscopic measurements}

As mentioned above, the essential physical predictions of CM are the truth and falsehood of the properties of a system. We will say that this information constitutes the \emph{outcomes} of microscopic measurements. Thus, given the state $x$, the truth function ${\mathcal T}_x(A)$ summarizes the outcomes of all microscopic measurements. It is a well-accepted feature of CM that there is no a priori limit on the number of measurements whose outcomes can be simultaneously determined. Furthermore, classical measurements can be arranged so that they have negligible effect on the state.

\vskip 10pt
\noindent
\underline{Determinism}

Classical mechanics is deterministic, in that \emph{if} the state is known exactly at some time, any property is exactly determined at any later (or earlier) time
to be either true or false. There are, however, two important qualifications to this determinism. The first is the well-known phenomenon of chaos, namely that in order to determine a property with an accuracy $\alpha$ at time $t>0$, one would in general have to know the state at time $t=0$ with an accuracy $\alpha \textrm{exp}(-\beta t)$, where $\beta$ is a time
constant characteristic of the system. Thus in practice the predictions are almost always approximate.

The second qualification is less often discussed but no less important. As alluded to above, classical mechanics does not tell you \lq what happens': all predictions of properties are contingent upon (\lq relative to', \lq \emph{contextual} to', or \lq \emph{conditioned} upon') an assumed state at some \lq initial time', but the theory does not tell us what that initial state is. By definition the state is declared to be \lq true' at some time (technically, the set containing only this state is \lq true')  and the predictions are the set of other true propositions concerning the truth of properties that follow from this assumption.  We shall see below that QM in its minimal formulation retains the conditioning with respect to the assumed truth of the initial state and it adds a conditioning with respect to frameworks (see Sec. \ref{3.1.3}).

\vskip 10pt
\subsection{Macroscopic theory (MACM)}\label{2.2}

The above is a complete formulation of what we call microscopic classical mechanics (MICM). We have not asked how states are prepared in the laboratory, nor how one might test whether the predictions are correct, i.e. whether they agree with experiment. In order to address those issues it is necessary to introduce preparation and measurement apparatus, which are necessarily macroscopic, so we are dealing with the second, or macroscopic, part of CM (MACM). The apparatus must be coupled to the \lq microscopic' system under study, which may itself be either small ($N = \textit{O}(1)$), or large ($N >> 1$), and in this way the predictions of MICM can be tested by MACM. From the above point of view it is seen that the macroscopic theory is a \emph{special case} of the microscopic theory, not a separate theory, if one considers the larger system which includes both the system $S$ and the apparatus as a new \lq microscopic' starting point.

Classical mechanics also illustrates a point concerning the terms \lq reality' and \lq real'. One could simply say that any element of the ontology is real, but this includes elements that are true as well as elements that are false, or elements that are indeterminate in case no initial state has been designated as true. We prefer to avoid the use of \lq real' and \lq reality' altogether in this context. We shall, however, sometimes use the term  \lq realistic' when referring to a \emph{formulation}, to which we give a well-defined meaning: a realistic formulation uses only concepts and language referring to the system under study, not external agents or apparatus. The microscopic theory is a realistic formulation.

\section{\label{3}	Quantum Mechanics}

 \vskip 10pt
Our formulation of QM will mirror the above treatment of CM as closely as possible, so that Question A of our Introduction is answered by a microscopic theory (MIQM) as in classical mechanics, with the macroscopic theory (MAQM) once again entering as an \emph{application} of MIQM to answer questions of experimental implementation and tests.  The essential difference between the classical and quantum cases results from the different basic ontology which in the quantum case is based on \emph{Hilbert space}. As in CM, our ontology will consist of states, which evolve in time, and properties, which do not (Schr{\"o}dinger picture). The essential complication arising in Hilbert space is that whereas in CM it is possible to make a consistent simultaneous assignment of truth values to \emph{all} properties, this is forbidden in QM as a consequence of quantum incompatibility (see below).

\vskip10pt
\subsection{Definitions}\label{3.1}
 \vskip 10pt
\subsubsection{ Hilbert space ontology}\label{3.1.1}
\vskip 10pt
We assume a system $S$ represented by an $N$-dimensional Hilbert space. For simplicity we shall often consider finite $N$, but the generalization to infinite $N$ presents no difficulties of principle.

\vskip 10pt
\noindent \underline{States}

Pure states are represented by rays of vectors, written in the Dirac notation as $\alpha |\psi \rangle$, where $\alpha$ is an arbitrary complex number of unit magnitude.  More generally, quantum states are represented by nonnegative operators of unit trace denoted as $\rho$, and often referred to as density matrices. For pure states these operators are \emph{projectors}, for which we use the notation $\rho = [\psi] =|\psi \rangle \langle \psi |$. Density operators which are not projectors, and therefore possess more than one eigenket, are known as \emph{mixed states}. [Different eigenkets of mixed states may be \emph{degenerate}, i.e. possess the same eigenvalue].

Observables are represented by Hermitian operators, of which the most important is the Hamiltonian $H$ which controls the deterministic time  evolution of states via the Schr{\"o}dinger equation.

\vskip 10pt
\noindent \underline{Properties}

Properties are represented by subspaces of Hilbert space, or equivalently by the corresponding projection operators onto those subspaces. The essential difference between CM and QM arises from the noncommutativity in general of two projectors. If two properties $A$ and $B$ have noncommuting projection operators
$[A]$ and $[B]$, we say that $A$ and $B$ are \emph{incompatible}. Following Kochen \cite{kochen2015} we shall refer to the set of all properties of a quantum system as its $\sigma$-complex, which we will denote as $Q(\mathcal{H})$.
\vskip 10pt
\subsubsection{Predictions}\label{3.1.2}
\vskip 10pt

 In the case of CM we saw in section II that given the state, the truth value of each property was determined uniquely via a truth function for pure states or a probability function for classical mixed states. An essential property of such functions is their treatment of logical operations (complementation, conjunction, disjunction), expressed through the truth tables \eqref{tfbool}, or their generalization to probability, the Kolmogorov conditions \eqref{kolminf} and \eqref{probmeas}, from which the overlap formula \eqref{kolm2} follows.

 For QM it is thus essential, if we wish to emulate CM, to define the logical operations as they apply to Hilbert space. In this way it was found by Birkhoff and von Neumann \cite{bvn} that quantum properties form a non-Boolean lattice. However, due to quantum incompatibility, the lattice members thus defined do \emph{not} obey the truth tables or the Kolmogorov relations. The main difficulty, discussed in \cite{fh1}, is the failure of the distributive law \eqref{clogdist} and its probabilistic counterpart \eqref{kolm2}, when applied to incompatible properties. Thus the lattice operations do not correspond to any legitimate truth or probability values when applied to incompatible properties.

 We shall therefore follow Kochen \cite{kochen2015}  and omit any reference to non-Boolean lattices. We shall only define the logical operations between \emph{compatible} properties $A$ and $B$ as

\begin{subequations}\label{qlog}\begin{eqnarray}
\text{NOT }A = \neg A = \text{orthogonal complement of }A, \\
A \text{ AND }B = A\wedge B = (A\cap B), \\
A \text{ OR }B = A\vee B = \text{span}(A, B),
\end{eqnarray}\end{subequations}

\noindent and note that the NOT ($\neg$) and OR ($\vee$) operations are different from the corresponding classical ones in \eqref{setlog}, since the orthogonal complement of a subspace $A$ contains only those vectors orthogonal to the vectors in $A$, and the span of two subspaces $A$ and $B$ contains all linear combinations of vectors in
$A$ and $B$, including those not belonging to either $A$ or $B$.

In order to generalize classical truth and probability functions to the quantum case we shall therefore need to identify appropriate sets of compatible properties which, following Griffiths \cite{grif1} , we call (property) \emph{frameworks}.

\vskip10pt

\subsubsection{Frameworks}\label{3.1.3}
\vskip10pt

Let us first introduce a \emph{sample space} $\mathcal{S}$ as a complete set of orthogonal properties

\begin{equation}
\label{sample1}
\mathcal {S} = \{A_i\}, \;\;\; A_i \perp A_j,\;\;\textrm{for}\;\; i \neq j,
\end{equation}

\noindent with $\vee_{i}A_i = \textrm{I}$,
or equivalently

\begin{equation}
\label{sample2}
\sum_{i}[A_i] = \textrm{I},\;\;\;\; [A_i][A_j] =[A_i] \delta_{i,j}\;(\textrm{no summation convention}),
\end{equation}

\noindent where I is the identity operator. This sample space is an orthogonal decomposition of the identity, which one can refer to as an Exhaustive Set of Exclusive Alternatives (ESEA). Note that the properties $A_i$ are not necessarily one dimensional.

Given a sample space $\mathcal{S}$ we form its \emph{algebraic closure} $\mathcal{F}$, by adjoining all properties obtained by complementation $\neg$ and countably infinite disjunction $\vee$ and including the empty property $\emptyset$, whose projection operator is 0. This closure is a Boolean algebra (even for countably infinite sample spaces) and it defines what we call the  \emph{framework} generated by $\mathcal{S}$.  .  It can be shown, however, that there is a one-to-one correspondence between sample spaces and their associated frameworks, since the members of $\mathcal{S}$ are precisely the \lq\lq atoms" of $\mathcal{F}$.  ($A\in \mathcal{F}$ is an atom of $\mathcal{F}$ if $A\neq\emptyset$ and $\mathcal{F}$ has no element $B$ such that $A>B>\emptyset)$.


By construction it is clear that all the properties in a framework are mutually compatible. Two frameworks are said to be compatible if all the properties of the first are compatible with all the properties of the second. Two compatible frameworks are not necessarily identical, but they share a common \emph{refinement}.


(A refinement of $\mathcal{F}$ is another framework $\mathcal{F}'$ that includes $\mathcal{F}$ as a proper subset.  Equivalently, $\mathcal{F}'$ contains, as events, all the atoms of $\mathcal{F}$, but not all of these are atoms in $\mathcal{F}'$: an atom $A$ of $\mathcal{F}$  may be replaced in $\mathcal{F}'$  by several "finer" atoms whose disjunction makes up $A$.)

\vskip10pt

\subsection{Truth functions and probability functions}\label{3.2}

\subsubsection{Quantum truth functions and the Bell-Kochen-Specker theorem}\label{3.2.1}
\vskip10pt

From the above discussion it is clear that no classical truth function can be defined over the $\sigma$-complex of all quantum properties of a system, for the simple reason that the logical operations are not defined for incompatible properties, so that the truth tables will not apply. We therefore define a \emph{quantum truth function} as a function assigning a truth value (T or F) to any property, but only require that the truth tables apply to \emph{compatible properties}.
Then a celebrated theorem, proved topologically by Bell \cite{bell1966}, for which Kochen and Specker \cite{koc} then provided an explicit construction, states that no such quantum truth function exists in any Hilbert space of dimension greater than 2.  (For dimension $>3$ a simpler counterexample has since been provided by Mermin \cite{Mer1990}.

\vskip10pt

\subsubsection{Quantum probability measures and Gleason's theorem}\label{3.2.2}

\vskip10pt

Given the nonexistence of quantum truth functions we turn to probability values, i.e. the probability that a property is true. Here again, no classical probability function can be defined over the full $\sigma$-complex, since the Kolmogorov relations cannot be satisfied for incompatible properties. We therefore define a \emph{quantum probability measure} $\cal{M}$ as a function assigning a value in the interval $[0,1]$ to each property $A$ in $Q(\mathcal{H})$, but again only required to satisfy the Kolmogorov relations \eqref{kolminf} and \eqref{probmeas} for compatible properties.

Given a quantum state $\rho$, a quantum probability measure $\mathcal{M_{\rho}}$ may be obtained from the relation

\begin{equation}
\label{Born1}
{\cal M_{\rho}}(A) = \mathrm{Tr}(\rho[A]),
\end{equation}

\noindent where $[A]$ is the projector associated with $A$ and $\mathrm{Tr}$ is the trace. This relation is of course known as the Born rule, but here we take it to define the quantum measure $\mathcal{M_{\rho}}$.

A powerful theorem due to Gleason \cite{glea} then shows that for a Hilbert space of dimension greater than 2, given a quantum probability measure $\mathcal{M}$, there exists a density operator $\rho$ such that $\mathcal{M} = \mathcal{M_{\rho}}$, as given by \eqref{Born1}. There is thus a one-to-one correspondence between states defined as density operators and as quantum probability measures. In fact, many authors (including Kochen) \emph{define} quantum states in terms of what we denote as quantum probability measures, which they are free to do given the correspondence.

The above arguments lend considerable strength to the presumption that the Born rule \eqref{Born1} is the \emph{only} consistent way a quantum state can assign a probability value to a quantum property. Indeed, as far as we know no one has proposed any different rule that would be consistent with all the features of Hilbert space. Nevertheless, our arguments do not constitute a rigorous proof, since we have certainly not shown that no other map, different from \eqref{Born1}, exists that would generate some other one-to-one correspondence between density matrices and quantum probability measures. It turns out, however, that by adding an extra assumption we can in fact \emph{prove} the uniqueness of the Born rule, as explained below. Let us first define noncontextuality.

\vskip10pt

\subsubsection{Contextual and noncontextual quantities}\label{3.2.3}

\vskip10pt
We will say that a quantity or statement is \emph{noncontextual} if it applies to a property independent of any framework containing that property. It is \emph{contextual} if the value of the quantity or validity of the statement does depend on the framework.

\vskip10pt
\noindent
\underline{Noncontextual network of probability functions}

Given a framework $\mathcal{F}$, the Born rule generates a unique classical probability function $P_{\rho,\mathcal{F}}$ whose sample space is the space $\cal{S}$ corresponding to that framework.  The value of $P_{\rho,\mathcal{F}}$ for any property $A\in \mathcal{F}$ is

\begin{equation}
\label{Born2}
P_{\rho,\, \cal {F}} (A) = {\cal M}_{\rho}(A) = \textrm{Tr}(\rho \, [A]).
\end{equation}

\noindent As already noted, $\mathcal{M}_{\rho}$ is not itself a probability function over the whole $\sigma$-complex $Q(\mathcal{H})$, since no such function exists. But if we let $\mathcal{F}$ range over all frameworks, we get a network of probability functions whose combined sample spaces do cover the whole $\sigma$-complex.  This network is noncontextual in that, for any fixed property $A$, \eqref{Born2} shows that $P_{\rho,\mathcal{F}}(A)$ has the same value Tr$(\rho [A])$ for every $\mathcal{F}$ that contains $A$.

The converse of this statement, namely that such a noncontextual network of probability functions can \emph{only} be realized by the Born Rule \eqref{Born2}, for some density operator $\rho$, is what we call the Noncontextual Network Theorem.  The proof of this converse statement is presented in Appendix B.
\vskip10pt
\subsection{Conditioning, registration and selection}\label{3.3}

An essential set of operations, which link the ontology of states and properties with the information-theoretic content of QM embodied in the truth of properties, are the operations of conditioning, registration and selection. Our treatment is largely based on the paper by Kochen \cite{kochen2015}, with some changes of language.

\vskip10pt

\subsubsection{Conditioning on a single property}\label{3.3.1}

\vskip10pt

In classical probability theory a probability function $P(x)$ conditioned on the event $y$ is defined by the equation

\begin{equation}
\label{condclas}
P(x|y) = P(x \wedge y)/P(y),
\end{equation}
\noindent where it is assumed that $P(y) \neq 0$. It is from the definition \eqref{condclas} that Bayes's rule is derived. Kochen then proposes the \emph{same} definition in the quantum case, for the probability value of $B$ given that $A$ is true in the state $\rho$:

\begin{equation}
\label{condqu}
P_{\rho}(B|A) = P_{\rho}(B \wedge A)/P_{\rho}(A),
\end{equation}
\noindent \emph{provided} $A$ and $B$ are compatible. Given the above definition, it follows that $P_{\rho}(B|A)$ is uniquely given by the relation

\begin{equation}
\label{cond3}
P_{\rho} (B|A) = \textrm{Tr} ([A]\rho[A][B])/\textrm{Tr}(\rho[A]).
\end{equation}

\noindent In view of the one-to-one correspondence between states and probability measures, we see that if the state $\rho$ is \emph{conditioned on the truth of the property $A$}, the resulting state is

\begin{equation}
\label{cond4}
\rho_A = [A] \rho [A] /\textrm{Tr}(\rho [A]),
\end{equation}

\noindent that is, for any $B$ we can write \eqref{cond3} as

\begin{equation} \label{cond4a}
   P_{\rho} (B|A) = \textrm{Tr}(\rho_A B),
\end{equation}

\noindent with $\rho_A$ given by \eqref{cond4}.

Although the above derivations involved only \emph{compatible} properties $A$ and $B$, if $\rho_A$ in \eqref{cond4} is a density matrix it must define probability functions $P_{\rho_A}$ via the Born rule for properties $C$ which are incompatible with $A$ as well, i.e. it must satisfy the relation

\begin{equation}
\label{cond5}
P_{\rho} (C|A) = P_{\rho_A}(C) = \textrm{Tr} (\rho_A C)),
\end{equation}

\noindent even when $A$ and $C$ are \emph{incompatible}

This last statement has also been proved by Kochen, though the proof as given is so concisely worded as to make it extremely difficult to follow all the steps.  (We present an expanded summary of his argument in Appendix C.)  Moreover, we note that \eqref{cond4} is the von Neumann-L\"{u}ders projection rule \cite{vn} for state reduction or \lq\lq collapse". As pointed out by Kochen, this rule here follows from our definitions and theorems about Hilbert space and does not require a separate axiom.

The definitions of conditional states given here illustrate an important underlying principle of QM emphasized by Preskill \cite{preskill}, namely that adding the information that the property $A$ is true to the state $\rho$, \emph{transforms} this state from $\rho$ to $\rho_A$. This means that information is an essential part of the ontology of QM. Preskill expresses this by the maxim \lq information is physical', but in our view the term \lq physical' is easily misinterpreted so we prefer to avoid it in this context.

\vskip10pt
\subsubsection{The law of alternatives}\label{3.3.2}
\vskip10pt

In classical probability theory the law of alternatives states that conditioning on the disjunction $y= y_1 \vee y_2$ of two events leads to the sum of the conditional probabilities

\begin{equation} \label{alt}
P(x|y) = P(x|y_1) P(y_1|y) + P(x|y_2) P(y_2|y).
\end{equation}

\noindent In the quantum case, on the other hand, \eqref{cond3} implies that if $A_1 \perp A_2$ and $A = A_1 \vee A_2$ then

\begin{equation} \label{altqu}
P_{\rho} ( B|A) = P_{\rho} ( B|A_1)P_{\rho} ( A_1|A)+P_{\rho} ( B|A_2)P_{\rho} ( A_2|A)+
  \textrm{Tr} ([A_1]\rho [A_2] [B]\!+\![A_2]\rho [A_1] [B])/\textrm{Tr}(\rho [A]).
\end{equation}

\noindent Thus in addition to the classical sum of conditional probabilities, the last term in \eqref{altqu} represents the effects of \emph{quantum interference}, which follows naturally from the definition \eqref{cond3} of conditional probability and once again does not require any additional assumptions.
\vskip10pt

\subsubsection{Conditioning on several properties: registration}\label{3.3.3}

Although \eqref{altqu} above in a sense represents conditioning on the two properties $A_1$ and $A_2$, it follows from the formula \eqref{cond3} for conditioning on the single disjunction $A = A_1 \vee A_2$. We now wish to consider quantum mechanical conditioning with \emph{registration} of each property. In MIQM this is \emph{defined} by the classical conditioning formula \eqref{alt}, i.e. the law of alternatives, with the interference term \emph{omitted}. For example, conditioning the state $\rho$ on a set $\{A_1, ..., A_K\}$ of \emph{disjoint} properties (i.e. $[A_i][A_j] = \delta_{ij} [A_i]$), with registration, yields the probability function

\begin{equation}\label{reg1}
P_{\rho}(B|A_1,...A_K) =\sum_{i}^{K}  P_{\rho}(B|A_i)P_{\rho}(A_i)/\textrm{Tr}(\rho \; [\bar{A}_K)],
\end{equation}
\noindent with
\begin{equation}\label{reg1a}
[\bar{A}_K] = \sum_{i}^{K} [A_i].
\end{equation}
\noindent The above probability function corresponds to the state
\begin{equation}\label{reg3}
\rho_K = \sum_{i}^{K}[A_i]\rho[A_i]/\textrm{Tr}( \rho\;[\bar{A}_K]).
\end{equation}

\noindent This simple result is obtained from \eqref{reg1} by using \eqref{cond3} for $P_{\rho}(B|A_i)$ and \eqref{Born2} for $P_{\rho}(A_i)$.

For the important special case where the set $\{A_i\}$ spans the whole Hilbert space it then forms a sample space $\mathcal{S}$  and the projector $[\bar{A}_K]$ is the identity operator, so \eqref{reg3} reduces to the mixed state

\begin{equation} \label{rhomix}
\rho_{\mathcal {S}} = \sum_{i} [A_i] \rho [A_i].
\end{equation}

\noindent

We may also say that the state $\rho$ is conditioned on the framework $\cal {F_S}$, which is uniquely determined (see Section 3.1.3) by the sample space $\cal {S}$, and write $\rho_{\cal {F}}$ in place of $\rho_{\cal {S}}$.

We shall discuss the important question of physical implementation of such registration below, but for the moment the microscopic definitions we have given are unambiguous.

\vskip10pt

\subsection{Subsystems and entanglement}\label{3.4}

The most interesting ways in which QM differs from CM are in the treatment of composite systems made up of different subsystems, e.g.  $S_1$ and  $S_2$, each one with its own Hilbert space $\mathcal{H}_1$ and $\mathcal{H}_2$, respectively. The Hilbert space $\cal{H}$ of the combined system $S = S_1 + S_2$ is then the \emph{tensor product} $\mathcal{H} = \mathcal{H}_1\otimes \mathcal{H}_2$ of the Hilbert spaces $\mathcal{H}_1$ and $\mathcal{H}_2$. The subject of composite systems is extremely rich and we shall only mention a few salient features. The reader is referred to Kochen \cite{kochen2015}, Preskill \cite{preskill}  and Bub and Pitowsky \cite{bp2}, among many other references, for further information.

 \vskip10pt
\subsubsection{Entanglement}\label{3.4.1}

We begin by defining a pure state of $\cal{H}$ which is \emph{not} entangled, i.e. a separable state, formed by taking the direct product of states in $\mathcal{H}_1$ and $\mathcal{H}_2$

\begin{equation} \label{sep}
|\Psi_{12}\rangle = |\Psi_1\rangle |\Psi_2\rangle .
\end{equation}

\noindent We may form the density matrix $\rho_{12}$ corresponding to the state \eqref{sep} and trace it over the states of either $\mathcal{H}_1$ or $\mathcal{H}_2$, to form the reduced density matrices $\rho_2$ and $\rho_1$, respectively. For a separable pure state of $\mathcal{H}$ such as \eqref{sep}, these reduced density matrices are pure states.

Any state of $\cal{H}$ which cannot be expressed as a direct product  
is \emph{entangled}. A simple entangled state is the pure triplet state consisting of two spins-1/2
\begin{equation} \label{sing}
|\Psi_s\rangle = \frac{1}{\sqrt{2}}(|\uparrow\rangle_1 |\uparrow \rangle_2+ |\downarrow\rangle_1 |\downarrow \rangle_2),
\end{equation}

\noindent where $|\uparrow \rangle $ represents the spin in the $+ z$ direction, say. The reduced density matrices $\rho_1$ and $\rho_2$ are now \emph{mixed states}, e.g.
\begin{equation} \label{mixed}
\rho_1 = \frac{1}{2} |\uparrow \rangle_1\; {_1\langle} \uparrow| + \frac{1}{2}|\downarrow \rangle_1\; {_1\langle} \downarrow|.
\end{equation}

It turns out that the same states \eqref{sing} and \eqref{mixed} may just as well be written in terms of the kets $|\rightarrow \rangle$, say, of spins polarized in the $x$-direction
\begin{equation} \label{mixedx}
\rho_1 = \frac{1}{2} |\rightarrow \rangle_1\; {_1\langle} \rightarrow| + \frac{1}{2}|\leftarrow \rangle_1\; {_1\langle} \leftarrow|,
\end{equation}

\noindent or any other direction for that matter. In the next subsection we point out that this symmetry is much more general.

A pure state of system $S_1$, e.g. $|\uparrow \rangle_1$ or $|\downarrow \rangle_1$ may be obtained from \eqref{sing} by conditioning on the $\{\uparrow, \downarrow\}$ framework of the system $S_2$ , no matter how far apart systems $S_1$ and $S_2$ might be.  This remote preparation process determines whether a state in the $\{\uparrow, \downarrow\}$ or the $\{\rightarrow, \leftarrow\}$ frameworks of \eqref{mixed} or \eqref{mixedx}, respectively, is thus produced, but of course not which of the two possible states in each case it will be. This phenomenon is known as \lq remote steering' and it was recognized long ago by Schr{\"o}dinger \cite{schrodinger1935,schrodinger1936}, though he expressed doubt about whether the effect could be observed experimentally.

\vskip10pt
\subsubsection{Mixed state ensembles}\label{3.4.2}

As mentioned above, the example in \eqref{mixed} can be generalized by considering the representation of the mixed state $\rho_{\mathcal{S}}$ in \eqref{rhomix} in terms of the states $[A_i]$. It is a well-known and essential feature of Hilbert space that this representation is also not unique. Indeed, as clearly explained by Preskill \cite[Sec. 2.5]{preskill}, there is a theorem due to Hughston, Josza and Wooters \cite{hughston}, to the effect that \emph{any} mixed state has a large number of ensemble representations in terms of different pure states, and no one of these representations is preferred. That is, there is no conditioning operation performed within the reduced Hilbert space of the mixed state \eqref{rhomix} which could distinguish between these representations.  Kochen, however, has argued \cite[Sec. 8.3]{kochen2015} that a preferred representation can be identified if information is available about the prior preparation of $\rho_{\mathcal{S}}$.

\vskip10pt
\subsubsection{No signaling and no cloning}\label{3.4.3}

We conclude this subsection by mentioning two important theorems which follow from the Hilbert space ontology. The first is a \emph{no cloning} condition according to which the arbitrary copying of an unknown quantum state is prohibited. As explained for example by Wooters and Zurek \cite{wz}, no cloning arises in QM because copying is a \emph{nonlinear} operation, whereas states evolve according to linear unitary transformations. There are many interesting consequences of the no cloning condition, only a few of which will be mentioned in what follows.

The no cloning condition is itself a consequence of the so-called no signaling principle (sometimes called Einstein locality), according to which instantaneous communication over long distances cannot be achieved by means of quantum operations. A cloning machine, however, would precisely permit such communication (see \cite{wz} or \cite{bp2}, among many other references).

\vskip10pt
\subsection{Quantum information}\label{3.5}

Strictly speaking all information is classical, i.e. in our usage information always  involves classical truth functions or their probabilistic generalization, classical probability functions. Thus, \lq quantum information' is a slight misnomer. It is shorthand for \lq the quantum flow of classical information', namely how information flows from states to properties in QM. We shall use the term quantum information, but we shall strive to remember what it stands for. In CM the state is the source of truth, so truth and falsehood flow instantaneously from states to properties.  In QM, on the other hand, there are no universal truth functions; so the flow of (classical) information is less straightforward.

\vskip10pt
\subsubsection{The breaking of framework symmetry}\label{3.5.1}

As we have seen, in QM the state is still the source of truth, but the state cannot transmit this information directly to properties. First of all the state can at most confer the probability of being true (or false) to a property, but even this is not directly possible in view of the nonexistence of a probability function over the whole $\sigma$-complex. The state can only confer probability within subsets of properties, i.e. within frameworks. As noted in Sec. \ref{3.2.3} the state defines a \emph{network} of probability functions, no one of which is singled out \emph{a priori}. In a slight abuse of language, we shall refer to this circumstance as \emph{framework symmetry}. It is only if this symmetry is \emph{broken} that a legitimate probability function can be defined over the selected framework. There is thus no direct connection between quantum states and properties, but on the other hand the state carries the information potentially for \emph{all} the frameworks, until this information can become \lq liberated' by the breaking of framework symmetry. This circumstance is at the origin of the power of quantum information.

\vskip10pt
\subsubsection{Logical time vs. dynamical time}\label{3.5.2}

So far we have made no reference to the dynamical evolution of the quantum state; so it is clear that the flow of quantum information brought about by the conditioning and selection operations can occur \emph{instantaneously} in dynamical time $t$. As mentioned earlier, such logical operations are examples of the all-important principle cited by Preskill, that information affects the quantum state. We will say that the flow of information occurs in \emph{logical} time, whose connection to dynamical time depends on the specific physical situation under consideration. Bub and Pitowsky \cite{bp2} refer to this flow as being a \emph{kinematic}, as opposed to dynamical process.

\vskip10pt

\subsection{Measurement in MIQM}\label{3.6}

In almost all formulations of QM, states are said to be prepared and measured, without precise definitions of these concepts. This was a principal complaint of John Bell [1] against textbook QM. Implicitly these actions are often thought of as being undertaken by agents, frequently called Alice and Bob, who in addition to preparing and measuring, are also sometimes said to look at the result and to hold beliefs.
In MIQM, we replace these notions by conditioning, registration and selection, which are all precisely defined mathematical operations pertaining to the system S alone. In Sec. \ref{3.6.1} we shall add the concept of \lq framework selection with interrogation' in our definition of microscopic measurement.

We need not refrain, even in MIQM, from using terms such as preparation and measurement, but in our minds they are always preceded by the word microscopic. In particular, microscopic Alice and microscopic Bob refer to conditioning and selection in one or another subsystem of a composite system. There are no external agents in MIQM. As explained below, however, our definition of microscopic measurement involves an implicit assumption which we wish to make explicit.

\subsubsection{The microscopic measurement outcome assumption}\label{3.6.1}

Microscopic measurements are in a sense at the heart of MIQM, so they must be defined with care. In Sec. \ref{2.1}
we defined microscopic classical measurements in terms of the truth function ${\cal T}_x(A)$, which determines the truth and falsehood of all the properties of the system. In QM truth functions are replaced by probability functions and they require selection of a state and a framework for their definition. These selections are necessary but not sufficient, since framework selection occurs in more general contexts than measurement, for example in Sec. \ref{3.3.3}, \eqref{rhomix} where it is used to prepare a mixed state. In order to define microscopic quantum measurements we introduce the concept of  \lq framework selection with interrogation'. The \emph{microscopic measurement outcomes assumption} is then stated as follows: a state $\rho$ is selected and a framework  $\cal F$ is selected \emph{with interrogation}. It is derived as in Sec. \ref{3.1.3} from a sample space
$\cal S$ and it defines a probability functionf whose domain is the properties of $F$. In addition, because of the interrogation condition, this selection also defines  a microscopic measurement which has a single outcome $A\in \cal S$. The theory does not determine which member of $\cal S$ is so designated, only its probability, given by the Born rule. Selecting this outcome produces a new state $\rho_A$ given by \eqref{cond4}.

We use the term \lq outcome' for members of the sample space, that is for atoms of the framework, and \lq event' for any member of the framework itself, that is any disjunction of some subset of the sample space.   It is possible to condition on any \emph{true} event $E\in\cal F$, that is on any event that is a disjunction of the single true outcome $A$ with other atoms. (Half of all the properties in the framework are true.) 	


Conditioning on a nonatomic true event is an act that arises naturally in various contexts.  As a minimal example, suppose that our system consists of two spin-1/2 particles and that $\rho = (I/2)_1 (I/2)_2$.  Choose the framework $\cal{F}$ whose atoms are the projectors
$[S_z^+]_1 [S_z^+]_2, [S_z^+]_1 [S_z^-]_2, [S_z^-]_1 [S_z^+]_2, [S_z^-]_1 [S_z^-]_2$. Say that the true atom is  $[S_z^+]_1 [S_z^+]_2$. Condition on the true event $E = [S_z^+]_1 = [S_z^+]_1 [S_z^+]_2 + [S_z^+]_1 [S_z^-]_2$. The resulting state is $\rho_E = (|S_z^+\rangle\langle S_z^+)_1 (I/2)_2$. This corresponds to the laboratory act of passing only the first particle through a Stern-Gerlach machine.  Of course there could be many more particles, and more complex (nonentangled) states. In general,we shall continue to speak of conditioning on $A$ except where we wish to generalize to nonatomic events\emph{}.

Although we have presented this definition of microscopic measurements in terms of an independent assumption, it is our view that it is the only consistent way in which measurements can be defined in MIQM. Note, moreover, the differences with MICM: (i) quantum measurements, as noted many times above, require the prior selection of a sample space $\cal S$, from which the outcome $A$ is selected; (ii) there is thus a severe restriction in the possibility of simultaneous quantum measurements: only measurements arising out of compatible frameworks can be made simultaneously; (iii) most important, 
.
the transformation from $\rho$ to $\rho_A$ may disturb the state.  We shall explore in Section 3.6.4 the condition for such disturbance.

\subsubsection{\label{3.6.2} Microscopic preparation}
 The microscopic preparation of a pure state is the selection of a vector $|\psi>$, which defines the projector $[\psi] = |\psi><\psi |$, a pure state
$\rho_{\psi} = [\psi]$. The microscopic preparation of a mixed state can be achieved by first selecting any state $\rho$ and then conditioning this state on a set $\{A_i\}$ of properties with registration, as in Sec. \ref{3.3.3}. The ensuing state $\rho_K$ , given by Eq. \eqref{reg3}, yields the desired microscopic preparation.

\subsubsection{\label{3.6.3} Free choice and intrinsic stochasticity}

As described above, preparation and measurement, defined in MIQM in terms of conditioning and selection with interrogation,
involve three different types of selection, namely of states, frameworks and outcomes (properties). We wish to note that the first two are subject to free choice, by which we simply mean that there is nothing in the theory which a priori restricts which state or framework may be selected. The theory merely examines the consequences of any such selection. The first such freedom is also present in classical mechanics; what QM adds is the second freedom. According to the microscopic measurement outcome assumption, however, the selection of outcomes is not subject to such freedom or control. (Here we specifically mean the selection of a member of $\cal S$.)

\subsubsection{\label{3.6.4}  Stochasticity and disturbance}

Let a state $\rho$ and a framework $\mathcal {F}$ be chosen (the latter with interrogation) in anticipation of a measurement.  Unless $\rho$ and $\cal F$ are compatible, there is no proceeding by which $\rho$ can be prepared so as to guarantee the production of a particular outcome when the measurement is performed; this applies even if $\rho$ is a pure state.  This fact, related to Heisenberg's uncertainty principle, is now widely regarded as a fundamental feature of quantum mechanics: QM is \emph{intrinsically stochastic}.  We regard this feature as stemming from the very structure of Hilbert space.

Stochasticity is often linked with the phrase ``irreducible and uncontrollable disturbance'', taken from \cite{bp2}.  Let us ask whether every microscopic measurement involves a disturbance.

The simplest example is that of a spin-1/2 particle in which $\rho = [S_x^+] = \frac{1}{2} + S_x$, the projection operator onto the ket $|S_x^+\!\!>$ which has positive $x$-spin.  If we were to ask how $\rho$ was prepared, it would be by interrogation of a prior framework $\mathcal{F}_x$ whose atoms are $[S_x^+]$ and $[S_x-^]$.  We choose a new framework $\mathcal {F}$ in order to generate a new state by interrogation.

If we choose the new framework to be $\mathcal {F}_z$, interrogation will yield either $[S_z^+]$ or $[S_z^-]$ with equal probability.  Only one of these will be the true state $\rho_A$; let us suppose it is $[S_z^+]$.  The passage from $\rho$ to $\rho_A$ involves both stochasticity, because the outcome was uncertain beforehand, and disturbance, because information is lost: it is not possible from the state $[S_z^+]$ to determine whether the prior state was $[S_x^+]$ or $[S_x^-]$.  But do stochasticity and disturbance always occur together?

We might have chosen $\mathcal {F} = \mathcal {F}_x$ so that the new framework is the same as the old.  In that case the outcome is certain to be $[S_x^+]$,
so that there is no stochasticity.  There is also no disturbance, since the measurement only confirms what was already true.  But let us look at more  examples.

Suppose that the initial state $\rho$ was not  $[S_x^+]$ but $I/2 = \frac{1}{2}([S_x^+] + [S_x^-)$, and  $\rho$ was prepared from the trivial framework
$\mathcal {F}_0 = \{I,0\}$. Let the new framework be $\mathcal {F} = \mathcal {F}_x$, and the true outcome again be  $\rho_A = [S_x^+]$.  Now there is stochasticity (the new state could not be predicted from the old state and the new framework) but no disturbance, because no information has been lost.  Before conditioning, only the event $[S_x^+] + [S_x^-]$ was determined to be true, and it is still true after conditioning.  Information has been added, not subtracted.

On the other hand, let us interchange the old framework and state with the new.  Now $\rho = [S_x^+]$, prepared from $\mathcal {F}_x$, and the new framework is $\mathcal {F}_0$.  The true outcome is $\rho_A = I/2$. Here there is no stochasticity, because $\rho_A$ (although a mixed state)
was completely determined as a density matrix by knowing $\rho$ and the new framework.  But there is disturbance, because $[S_x^+]$ is no longer determined to be true after conditioning; information has been lost.  (This operation should perhaps be called state preparation, rather than measurement as described in Section 3.6.1.)

Thus we have stochasticity without disturbance in one direction, and disturbance without stochasticity in the other.  The effect is not dependent on the triviality of the framework $\mathcal {F}_0$: we may consider a spin-1 particle with states $[S_x^+]$, $[S_x^0]$, $[S_x^-]$, and instead of the trivial framework we may consider one whose atoms are $[(S_x^2)^+] = [S_x^+] + [S_x^-]$ and $[(S_x^2)^0] = [S_x^0]$.  Then the nontrivial event $[(S_x^2)^+]$ plays the same role that $I$ played in the previous example, and the reasoning goes through as before.

An examination of these and several other examples suggests the following principle: that \emph{disturbance occurs in a microscopic measurement in which the new framework used for conditioning lacks one or more events that were present in the old one.} That is, the new framework, while differing from the old framework, is not merely a refinement of it (see Section 3.1.3).  We shall refer to this principle as the \emph{nonrefinement criterion} for disturbance.  We note that the term \lq\lq sample space" may be substituted for \lq\lq framework" in this criterion without changing its logical consequences.

It is worth comparing the nonrefinement criterion with the discussions of "disturbance" found in Bub(\cite{bub}) and in Preskill(\cite{preskill}), although both of these authors are discussing real measurements in the laboratory rather than microscopic measurements as we have defined them in MIQM.

Bub speaks of \lq\lq observer beliefs" but his ideas are easily restated in terms of frameworks.  He says (Section 6, pp. 250-2) that mere
\lq refinement'  is \lq\lq gentle" and does not constitute a disturbance, only \lq\lq readjustment", which is \lq\lq violent", does. He gives an example of the latter in which, as a consequence of a measurement, an outcome previously ruled out (having zero probability) becomes possible.

Preskill says that disturbance takes place whenever a procedure \lq\lq distinguishes between nonorthogonal states".  Observe that what we have called a refinement - passage from an old framework to a new one in which some former atoms, while retained as events, are split into new atoms representing finer alternatives - does not meet this citerion for disturbance, as the new atoms are still mutually orthogonal.

Allowing for differences in language, it appears that the spirit of these authors is closely represented by our nonrefinement principle, which identifies \lq\lq disturbance" with the passage from an old framework to a new one which does \emph{not} contain every property belonging to the old one.  In other words, certain possibilities of distinction, present in the old, are lost in the new.

\subsection{\label{3.7} The measurement problem}

The standard statement of the measurement problem refers to the collapse of the wave function upon measurement, one of the basic assumptions of orthodox (textbook) QM. This collapse, seen as a transformation of the quantum state, is in conflict with the unitary evolution of the quantum state implied by the Hamiltonian, whence the \lq problem'.

Although we have not based our definitions on (macroscopic) measurement, a version of the problem can be stated in MIQM as well. It is the fact that the von Neumann-L\"{u}ders projection rule \eqref{cond4}, a transformation  $\rho\rightarrow \rho_A$ which governs microscopic measurements (see Sec. \ref{3.6.1}), also violates the unitary evolution of the quantum state.  (We note that the von Neumann-L\"{u}ders ``collapse'', if not trivial, satisfies our condition for "disturbance" if the new framework governing the passage from $\rho$ to $\rho_A$ is compared with an old framework from which $\rho$ was generated.)

As discussed in Sec. \ref{3.5} above, it is the very structure of Hilbert space which prevents the direct flow of information from $\rho$ to $\rho_A$, due to the absence of universal truth functions. The projection rule, which follows directly from the Born rule \eqref{Born2}, is the only means by which information (truth) can be transferred from the state to any quantum property. Moreover, as emphasized above, the transformation occurs in logical time as opposed to dynamical time, so the question of consistency with the unitary dynamical evolution does not arise. Thus our resolution of this problem in MIQM is to declare that it is neither about (macroscopic) measurement, nor is it a problem: it is inherent in the very structure of Hilbert space and in how information can be transmitted by quantum states. We prefer to refer to this phenomenon simply as the the flow of quantum information, or at most the paradox of quantum information. In the next section we shall comment briefly on the relationship between our resolution of the microscopic measurement problem and its implementation through macroscopic measurements.

\vskip10pt
\subsection{\label{3.8} Two-slit interference}

The different conditioning events described above are well illustrated by the example of two-slit interference, which Richard Feynman \cite{feyn1} called  `` the \emph{heart} of quantum mechanics, ... the \emph{only} mystery''.  In our discussion of two-slit interference we will at times use the traditional language of preparation and detection, but since we are still talking about MIQM it should be interpreted as \lq microscopic' preparation and detection, i.e. in terms of conditioning, registration and selection. Let us assume a pure state $|\psi\rangle$, or its density matrix $[\psi]$, in the form of a plane wave prepared at time $t_0$ and incident on a double slit system at time $t_1$. Let the property of going through slit $i$ be $A_i, (i = 1,2)$. Then if the state $[\psi]$ is allowed to propagate through the slit screen without registering which slit the wave went through, this corresponds to preparing the pure state

\begin{equation}\label{2slitpure}
[\psi^{\prime}] = [A] [\psi] [A] / \textrm{Tr} ([\psi] [A]), 	
\end{equation}		
\noindent where  $[A] = |A\rangle \langle A|$ and
\begin{equation}\label{2slitstate}
|A\rangle = (1/\sqrt{2}) ( |A_1\rangle + |A_2\rangle),
\end{equation}		
\noindent is the pure state superposition of the vectors $|A_1\rangle$ and $|A_2\rangle$. Since all the cases where the incident wave $|\psi\rangle$ did not go through either one of the slits have been discarded, the state $[\psi^{\prime}]$ must be renormalized as in \eqref{2slitpure}. A new state has been prepared and after propagation to a detection plane it will predict the tell-tale \emph{interference pattern} of two-slit diffraction.

If, on the other hand, the state $[\psi]$ is conditioned on $A_1$ and $A_2$ at $t_1$, \emph{with} registration of which slit the wave went through but no selection of either property, then this predicts a mixed state as in \eqref{reg1} above with $i =1,2$, i.e.
\begin{equation}\label{2slitmixed}
\rho^{\prime} = ([A_1] [\psi] [A_1]+ [A_2] [\psi] [A_2])/ \textrm{Tr}( ([A_1] + [A_2]) [\psi]).	
\end{equation}

\noindent We could also condition on only one of the properties $A_1$ or $A_2$, with selection, thus preparing a pure state which would after propagation, predict a single-slit diffraction pattern at the detection plane with no interference. The mixed state \eqref{2slitmixed} on the other hand, where both $A_1$ and $A_2$ are registered but neither is selected, predicts a superposition of two non-interfering single-slit diffraction patterns at the detection plane.

As mentioned earlier, the above example illustrates what we call \lq microscopic preparation and measurement' of a quantum state, defined entirely in terms of (logical) operations pertaining to the microscopic system under study, as discussed in Subsection 3.6 above.

 \section{\label{4}	Macroscopic Quantum Mechanics (MAQM)}

The preceding section was our relatively brief summary of MIQM, a theory which, like its classical counterpart MICM, we claim to be logically complete. As was true in the classical case, however, a complete treatment must describe how the microscopic theory relates to experiments, i.e. how it can be implemented in the laboratory or how it can serve as a description of the natural world. In particular, the operations of state preparation and measurement must have an implementation which involves macroscopic experimental apparatus.

The first point to make is that although we have emphasized the contrast between CM and QM and between the microscopic and macroscopic theories, the most general formulation is simply MIQM, since it encompasses everything: classical mechanics is a special case of MIQM, as are systems (either classical or quantum), containing macroscopic apparatus.

That CM is a special case of QM can be \lq proven' in a variety of ways, with a range of levels of rigor, but it is a given in all our discussions. The classical limit involves large systems ($N \rightarrow \infty$), and consideration of macroscopic states and properties, though these should be considered as necessary rather than sufficient conditions for achieving the classical limit. The details of the limiting procedure and the quasiclassical realm through which the limit is attained are fascinating subjects in their own right, but since we have nothing to add to the many standard treatments, we will bypass this discussion.

In order to see how macroscopic measurements fit into MIQM, let us consider a large closed system $S^\prime$ consisting of the (small or large) \lq quantum' system $S$ under study, a classical measuring apparatus $\cal{M}$ and an environment $\cal{E}$ which is itself macroscopic, $S^\prime = S + \cal{M} + \cal{E}$. In the present section when we refer to the \lq microscopic or quantum system' we shall mean $S$, even though as noted above, both $S$ and $S'$ are quantum and described by MIQM.

Since we have already \lq solved' the measurement problem for the system $S$, we only need to show how this solution can be implemented in the laboratory, i.e. how it transfers to $S'$. We need to explain how the selected \lq true' property $A_k$, say, of $S$ is related to the \lq pointer reading' recorded by $\cal{M}$. Our treatment follows  the standard quantum measurement theory initiated by von Neumann \cite{vn}, as supplemented in more modern treatments with the phenomenon of decoherence introduced by Zeh \cite{z21} and elaborated for example by Zurek \cite{z9}, or Gell-Mann and Hartle \cite{gh3}, among many others.

The physical interaction of the measuring apparatus $\cal{M}$ with the system $S$ has two important functions: first, it provides a \emph{framework selection} mechanism for the measured system $S$. The simplest example is to consider the measured system $S$ to be a spin-$\frac{1}{2}$ in a state $\rho$, say, and the apparatus a Stern-Gerlach magnet. Orienting the magnet in the $x$-direction on the Bloch sphere \emph{selects} the $\{[S_x^+], [S_x^-]\}$ framework of the system $S$. Then according to MIQM applied to the system $S$, the combination of a state $\rho$ and a selected framework leads to a probability function, for which we assume the single microscopic  outcome, $[S_x^+]$, say. The second consequence of the interaction between $S$ and $\cal{M}$ is to \emph{entangle} the properties of the selected framework of $S$ with the properties of $\cal{M}$, which are the above mentioned pointer readings. These latter properties are, however, also in contact with the environment, whose role is to cause the properties of $\cal{M}$ to \emph{decohere}, i.e. to lose any mutual correlation. The one property of $\cal{M}$ which is correlated with the selected outcome in $S$ then becomes the \lq true' reading or \lq macroscopic measurement outcome', which can then be \emph{recorded}.

In this treatment, moreover, no additional measurement problem arises from the incorporation of $S$ into $S'$, since it has already been \lq resolved' within $S$. Nevertheless, the knowledgeable reader will recognize that the above description is oversimplified, since it omits the fact that the interaction between $S$ and $\cal{M}$ which is responsible for the logical process of framework selection in $S$, is a process which occurs in dynamical time in the experiment. In this case, although there remains a distinction between logical and dynamical time, the logical process of framework selection does not occur at a single instant of physical time. There is nevertheless no conflict between the logical and dynamical processes; the dynamical interaction between $S$ and $\cal{M}$ simply brings about the choice of framework and selection of outcome which are both logical processes within $S$.

The free choice mentioned in Sec. \ref{3.6.2} is manifested in MAQM by the statement that the experimenter is a priori free to prepare any state and to measure any observable. The theory places no general restrictions on these choices or selections, even though, of course, practical restrictions arising from the specific experiment under consideration may restrict the choices. The selection of a measurement outcome (pointer reading), on the other hand, is intrinsically stochastic, a fact that is well accepted in MAQM. What we have pointed out is that the same distinction between free choice and intrinsic stochasticity is already present in MIQM.

\section{\label{5}	Other Formulations}

In our previous publication FH1, we discussed other formulations and interpretations in some detail, albeit from the point of view of the histories formulation which we have not adopted in the present work. Nevertheless, the discussion of prior treatments in FH1 is largely independent of the histories point of view, so we shall present an abbreviated discussion here.

\vskip10pt

\subsection{\label{5.1} Traditional formulations}

\subsubsection{\label{5.1.1} Copenhagen operationalism}

The generally accepted \lq interpretation' of QM is the operationalist formulation of Niels Bohr, usually referred to as \lq Copenhagen'. The existence of a classical domain governing measurement apparatus is posited from the outset and all predictions must refer back to classical measurement outcomes. The Born rule is a basic postulate of the theory and it refers to the probability of such measurement outcomes. Excellent expositions of Copenhagen QM are in the textbooks of Landau and Lifshitz\cite{ll6} and Peres \cite{peres}. We consider this point of view to be an appropriate \emph{phenomenology}, but as emphasized by John Bell \cite{b7}, it is hard to accept as a complete theory. It immediately leads to the question \lq What is it a phenomenology of?'. Bohr, however, was adamant that no further elucidation of the underlying mechanisms behind the rules of QM was possible.

There are modern operationalist formulations that replace measurement apparatus with external \lq information agents', who serve to record or prepare quantum systems \cite{bz}, \cite{qbism}, \cite{fuchsetal}. In all of these formulations, including Copenhagen, the quantum state is merely an auxiliary device by means of which predictions can be made, but no quantum ontology is recognized.

\vskip10pt

\subsubsection{\label{5.1.2} Textbook or orthodox QM}

It is useful to distinguish, as we did in FH1, a philosophically less rigid \lq standard' formulation, originated by Dirac \cite{dirac} and pursued by von Neumann \cite{vn}, and subsequently by many, if not most, of the best QM textbooks. There, the properties of Hilbert space are central, and once again the Born rule is a basic axiom, stated in terms of the undefined concept of \lq measurement'. This, as mentioned above, was the main complaint of John Bell \cite{b7} against textbook formulations of QM. A second basic axiom is the von Neumann-L{\"u}ders projection rule \eqref{cond4}. Once these rules have been postulated, however, the subsequent developments are rather similar to our own discussion.

\vskip10pt

\subsubsection{\label{5.1.3} Modern Hilbert space treatments}

The similarity of textbook QM to our minimal formulation is particularly apparent in a number of modern treatments that are firmly based on Hilbert space. We have already singled out the recent work of Kochen\cite{kochen2015} which had a decisive influence on our views, as well as the paper of Bub and Pitowsky \cite{bp2} and the impressive lecture notes on quantum information by Preskill \cite{preskill}. Although these authors still use the traditional terminology of \lq measurement', the arguments are all essentially microscopic, so one can reinterpret their language by changing \lq measurement' to \lq microscopic measurement' and by understanding the actions of \lq Bob' and \lq Alice' to mean conditioning and selection operations within appropriate subspaces of Hilbert space. The subsequent developments then resemble our own rather closely.

We can illustrate the above statements by first considering the abstract of Bub and Pitowsky \cite{bp2}, in which the authors assert that in order to solve the measurement problem one must \emph{reject} two traditional dogmas of QM. The first is John Bell's assertion that  ``measurement should never be introduced as a primitive process in a fundamental mechanical theory like classical or quantum mechanics, but should always be open to a complete [dynamical] ... analysis." Our attitude toward this assertion is not complete rejection, since we agree with Bell that the ill-defined term \lq measurement' should not be part of the basic formulation. On the other hand, we do agree with Bub and Pitowsky regarding the second part of the quote, namely that a complete dynamical analysis is not necessary. The second dogma that Bub and Pitowsky reject is the ``view that the quantum state has an ontological significance analogous to the significance of the classical state as the \lq truthmaker' for propositions about the occurrence and non-occurrence of events, i.e. that the quantum state is a representation of physical reality." Here again, we accept one part of the statement, namely that the quantum state has an ontological significance, but of course not as a truthmaker for propositions, since there are no truthmakers in Hilbert space. Moreover, the last part of the statement, referring to physical reality, is not precisely defined and in our opinion is best left out of the basic formulation. Thus, while we largely agree with the message of the abstract, we find the language open to misinterpretation in many places. The substantive closeness  of our formulation to that of Bub and Pitowsky becomes clear, however, if one considers their Section 2, which provides a summary, entitled \lq An Information-Theoretic Interpretation of Quantum Mechanics'. The entire formulation is closely tied to Hilbert space, emphasizing as it does the absence of truth functions and the central role of Gleason's theorem in justifying the Born rule. As mentioned earlier, our distinction between logical and dynamical time is expressed by Bub and Pitowsky as a distinction between kinematic constraints and the dynamical evolution of the quantum state. However, these authors give a more prominent place to the principles of no signaling and no cloning, since they assign to them a motivational significance, as a justification for basing the theory on Hilbert space, in analogy to the role of Einstein's speed-of-light postulate in relativity. While we consider this point of view interesting, our formulation simply posits Hilbert space and then explores the consequences of this fundamental axiom. In contrast to Bub and Pitowsky, we do not seek \lq justification' for Hilbert space. Other than these differences of presentation, however, it is clear from Bub and Pitowsky's summary that their formulation is very close in all its essentials to ours.

Turning to Kochen \cite{kochen2015}, the similarity with our formulation is manifest throughout his paper. The main apparent difference is that in place of our distinction between the microscopic and macroscopic theories, Kochen bases his formulation on the distinction between \lq intrinsic' and \lq extrinsic' properties. Classical properties pre-exist any measurement and they are intrinsic. Quantum properties, on the other hand, are \lq relational' or \lq extrinsic' to appropriate measurements. In his Introduction Kochen states that this notion is expressed mathematically by saying that every experiment yields a $\sigma$-algebra of measured (extrinsic) properties and that the set of all quantum properties, the $\sigma$-complex, consists of the union of all the $\sigma$-algebras elicited by different decoherent interactions, such as measurements. We may re-interpret Kochen's description within MIQM by the statement that while classical states lead directly to a set of true and false properties, in QM it is necessary to specify a framework ($\sigma$-algebra), before truth and falsehood, or the probability thereof, can be determined. Since Kochen did not make our distinction between MIQM and MAQM, his description can also be viewed, perhaps more naturally, as anticipating MAQM.

Finally, let us comment on the lecture notes of Preskill \cite{preskill}, which are particularly sparing of interpretation. Their emphasis on Hilbert space and its multiple physical consequences, makes the whole discussion very similar to ours, with of course much more detail. The main difference between Preskill's formulation and ours is his introduction of the Born rule as a primitive axiom about the probability of \lq outcomes of measurements', a term which is not further defined. As mentioned above, however, we may interpret the word in terms of conditioning and selection , i.e. what we have termed \lq microscopic measurement'. Preskill's subsequent treatment, including his reference to Gleason's theorem, is then not fundamentally different from ours. Indeed, his Summary presented in Section 2.7, provides a microscopic  definition of a measurement, as an \lq orthogonal projection'. We have already mentioned Preskill's reference to the fundamental principle of QM that \lq information is physical', which we endorse fully, though we prefer to avoid the ill-defined term physical in this context.

The above remarks are intended to clarify our statement that the modern treatments we have singled out are, apart from language, not fundamentally different from our minimal formulation.

\vskip10pt

\subsection{\label{5.2} Many-worlds formulation}

We have already commented on this theory in FH1, so we will only add a few comments. We first note that the original motivation of Everett \cite{ev1} was the impossibility of formulating an operationalist quantum cosmology, since such a theory would require \emph{external} devices or agents. Our response to this observation is first that many subsequent formulations of QM go beyond operationalism, thus eliminating the main motivation for the many-worlds theory. Most important, in our opinion, is the additional point that formulating a cosmological theory should not be the basic starting point of mechanics, be it classical or quantum, but rather an \emph{application} to the whole universe, of the general microscopic theory, which is valid for any closed system of arbitrary size $N$. As mentioned above, such an application is always possible once the need for external devices or agents has been eliminated. We will not discuss the many-worlds point of view further, referring the reader to FH1 and to many other critiques and discussions.

\vskip10pt

\subsection{\label{5.3} Histories formulations}
\vskip10pt
\subsubsection{\label{5.3.1} Consistent histories}

The first of the histories formulations, the so-called \lq consistent histories' theory, is due to Griffiths \cite{grif2,grif1}, whose main motivations were (i) to provide a theory that did not rely on measurement in its basic axioms, and (ii) to formulate a quantum theory of probability that was consistent with the (Kolmogorov) axioms of probability theory. These requirements led Griffiths to base his formulation on so-called \lq histories' and \lq history frameworks', which carried all the essential information and predictions of QM. A similar formulation was presented some years after Griffiths's original work by Omn{`e}s \cite{omnes1992} and by Gell-Mann and Hartle \cite{gh3}. Our earlier work, FH1, was entirely devoted to a reformulation of the histories theory, with the aim of justifying some of its assumptions by arguments based on a more detailed exploration of Hilbert space.

FH1 divided the theory into two distinct parts, a \lq static theory', comprising two-time histories, which turn out to be always \lq consistent', and a \lq dynamic theory' comprising multi-time histories which require additional \lq consistency conditions'. From the point of view of the present work, we can say that the static theory of FH1 is essentially the same as our minimal formulation. Indeed, the two times of the static theory are a \lq preparation time $t_0$' and a \lq measuring time $t_1$', both of which are also defined in our formulation. In the present work we employ the Schr{\"o}dinger picture in which it is the state that evolves in (dynamical) time $t$ and properties have no time dependence. Thus, having prepared the state at $t_0$, we let it evolve under the Hamiltonian to $t_1$, at which point the \emph{microscopic} measurement operations of conditioning and selection can take place. This is how the present work connects to the static histories theory. [Actually, we might as well set $t_0 = t_1$, since the evolution of the state plays no role in the presentation]. Thus, the essential content of our formulation retains the fundamental ideas of sample spaces, frameworks and the single-framework rule first introduced by Griffiths.

The main difference between that part of the treatment of FH1 and the present work is our explicit use of conditioning, registration and selection in the latter, influenced as we were by the elegant \lq reconstruction' paper of Kochen  \cite{kochen2015}. Our formulation is thus not substantively different from the static part of FH1, but the presentation is more elegant and succinct. In particular, much of the content of Appendix B of the earlier work could be short-circuited by applying algebraic concepts described by Kochen.

The main advantage of following Kochen's use of conditioning and selection, however, is that it allowed us to understand that the whole dynamical theory of multi-time histories and history frameworks can be discarded, since most physical questions of interest can be addressed using the Schr{\"o}dinger picture for the dynamical evolution of states and applying the logical operations of conditioning and selection at fixed time (static theory). In this way the whole difficult question of consistency conditions can be bypassed. The necessity of imposing these conditions arises, in our opinion, from the very structure of multi-time histories: they involve repeated conditioning operations at a sequence of times, thus \emph{mixing} logical and dynamical time in their very definition. As noted early on by Griffiths, many multi-time histories thus defined turn out to be \emph{intrinsically inconsistent}, so that they need to be banned from the theory by hand, so to speak, in order to achieve consistency. Restricting discussion to the static theory, as we do in the present work, bypasses this troublesome circumstance.

\vskip10pt
\subsubsection{\label{5.3.2} Decoherent histories}

The decoherent histories theory, introduced shortly after Griffiths's initial paper by Gell-Mann and Hartle \cite{gh3}, had as its main motivations (i) to formulate quantum cosmology in a less obscure manner than in the many-worlds theory, and (ii) to provide a convenient formalism for describing the classical limit of QM, in particular the \lq quasiclassical realm' by which the classical limit is reached. The theory is mathematically equivalent to consistent histories, but the emphasis is on coarse-graining operations applied to sets of microscopic histories, which eventually lead to classical orbits in phase space if pushed to their limit. Multi-time histories are defined from the outset and the consistency conditions are applied in the quasiclassical realm, where they may be interpreted in terms of the physical phenomenon of decoherence, whence the name the authors chose for their formulation. Our comments regarding cosmology as an inappropriate \emph{starting point} for formulating QM are also applicable here, though the authors have been successful in largely overcoming the other obscurities of the many-worlds theories. Our comments concerning the basic difficulties associated with multi-time histories, on the other hand, certainly apply to decoherent histories. We are not saying that for certain specific purposes, e.g. study of the classical limit of QM, consideration of coarse grained multi-time histories is uninteresting. Our point is that this is an example of an application of the basic theory, not a necessary part of its foundations.

\vskip10pt
\subsection{\label{5.4} Neoclassical theory: the de Broglie-Bohm formulation}

As is well known the theory was first presented by de Broglie \cite{debroglie} at the dawn of QM, but he was so strongly criticized that he abandoned the effort, and it was not until David Bohm rediscovered the theory \cite{bohm2}, that de Broglie himself took it up once more. Here again, we refer the reader to our discussion in FH1 and make only a few additional comments. We have called the theory \lq neo-classical' because the basic ontology consists of the classical coordinates (i.e. configuration space rather than phase space) and the guiding wave function operates in a classical manner. Hilbert space has no fundamental physical significance; it is at most a formal construct governing the time dependence of the guiding field $\psi$. The primary motivations for both de Broglie and Bohm were first to replace the operationalism of Copenhagen with an ontology which would not depend on ill-defined concepts for its definition, a point of view with which we clearly sympathise. A second motivation was of course to solve the measurement problem. In our view, however, the theory - however ingenious - is profoundly misguided, since its classical ontology misses the essential physical elements of QM, which derive from Hilbert space and lend quantum processes and quantum information their unique and \lq miraculous' features. It is thus not surprising to us that the theory has attracted relatively little interest in the physics community.

\vskip10pt

\subsection{\label{5.5} Spontaneous collapse theories}

In contrast to all the other formulations and interpretations we have discussed, the spontaneous collapse theories, e.g. \cite{ghirardi}, solve the measurement problem by \emph{changing} QM. Specifically these theories introduce a physical collapse mechanism for the wave function through a stochastic force added to the Schr{\"o}dinger equation, whose strength is adjusted to be unobservable directly, but to indeed collapse the wave. Since the essential motivation for the theories is to solve the measurement problem, which we do not consider to be a problem within Hilbert space QM, we see little value in this effort. Of course, since the theories modify QM, they are potentially verifiable or falsifiable experimentally, but until such tests are actually forthcoming, we consider the theories to be of limited interest.
\vskip10pt

\subsection{\label{5.6} Excess baggage}

This term was introduced in the present context by Hartle \cite{excess} and in our view it characterizes most of the theories we have described in this section, to a greater or lesser extent. Our primary aim, to present the \emph{minimal} formulation of QM, was precisely to discard as much of the huge quantity of material comprising the literature on \lq foundations of QM' as possible. Our efforts include, in particular, removing macroscopic measurement apparatus or external agents which occur in operationalist formulations, discarding multi-time histories as discussed above, and setting aside all neo-classical approaches. It is not that we claim any or all of these notions to be \emph{incorrect}, merely that they are unnecessary to provide a clear and unambiguous answer to our fundamental question \lq What is QM?', or to fulfill John Bell's challenge expressed in his last lecture \cite{b7}:

\begin{quote}
Surely, after 62 years, we should have an exact formulation of some serious part of
quantum mechanics? By \lq\lq exact" I do not of course mean \lq\lq exactly true". I mean only that the
theory should be fully formulated in mathematical terms, with nothing left to the discretion
of the theoretical physicist,... until workable approximations are needed in applications... Is it not good to know
what follows from what, even if it is not really necessary for all practical purposes (\lq FAPP')?
\end{quote}

\section{\label{6}	Summary and Conclusion}

In this section we summarize the main elements of our minimal formulation of QM.

\begin{itemize}

\item The formulation uses as its template microscopic classical mechanics (MICM), whose ontology represents closed systems by objects in Euclidean phase space. Each closed system $S$ with $6N$ degrees of freedom, has a unique state represented by a point or a set of points in phase space. States evolve in dynamical time $t$ according to the Hamiltonian. Properties of the system are represented by subsets of phase space.
\item In CM states confer truth or falsehood on properties. The predictions of the theory are the set of true properties at any time $t$ and there is no limit on the number of properties that can be simultaneously true.
\item In the quantum case (MIQM) Euclidean phase space is replaced by Hilbert space, but the logical development is otherwise similar.
\item The ontology of MIQM consists of pure and mixed states represented by rays of vectors or more general density operators, respectively, and properties are subspaces of Hilbert space.
\item The essential feature of QM is quantum incompatibility, which results from the noncommutativity of projectors representing incompatible properties.
\item Two essential theorems of Hilbert space, the Bell-Kochen-Specker theorem and Gleason's theorem, demonstrate that in contrast to the classical case, no consistent assignment of truth values can be made for the full set of properties of a quantum system. Instead, states confer the probability of being true on properties, once a consistent sample space, called a framework, has been identified.
\item A given state defines a network of potential probability functions, each one associated with a different framework. A consistent probability functions is thus only defined once the state has been selected \emph{and} the symmetry of frameworks has been broken. Probability functions are thus conditional upon, or contextual to, the selection of both a state and a framework.
\item As explained in Sec. 3, the essential logical operations in MIQM are selection of states, frameworks and properties, as well as conditioning on properties and frameworks.  There is an important additional operation, namely 'framework selection with interrogation".  Once a probability function has been identified by selecting a state and a framework with interrogation, a single atomic property is the true outcome, though the theory does not specify which member of the sample space it is. Only the probability of that outcome is determined by the theory via the Born rule. This is what is meant by the statement that measurement outcomes are intrinsically stochastic.  An exception can occur when the old state is compatible with the framework; then (see Sec. 3.6.4) the outcome is predictable with certainty. (We also argue in the same section that \lq\lq disturbance" is not always coincident with stochasticity.)  In contrast, the preparation of states and the selection of frameworks are subject to free choice, since the theory contains no a priori restrictions of principle for such selections. The same freedom is also present in CM in the preparation of states.
\item The operations of conditioning and selection of states and properties occur in logical time, which is distinct from the dynamical time $t$. Logical operations can (but need not) occur simultaneously with respect to the time $t$.
\item Quantum states can change either due to logical processes ruled by logical time, or through dynamical processes evolving in dynamical time. This dual mode of change is what resolves the so-called \lq measurement problem' of consistency between the collapse of probability functions to select outcomes, and the unitary evolution of the quantum state in dynamical time.
\item In both classical and quantum mechanics the applicability of the theory to experiment is treated in a macroscopic theory (MACM or MAQM), which is a special case, or application, of the more general microscopic theory to describe, for example, state preparation or the measurement of properties. Once the measurement problem has been resolved in MIQM there are no further conceptual difficulties in understanding macroscopic quantum measurements.
\item The free choice of states and frameworks mentioned above is manifested in MAQM by the statement that the experimenter is a priori free to prepare any state and to measure any observable. The specific outcome of the measurement, however (the true pointer reading), is intrinsically stochastic. This distinction is implicit, if not explicitly stated, in standard presentations  of textbook QM, but we have deduced it from a similar distinction which exists already in MIQM.
\end{itemize}

We conclude by returning to the four questions in the Introduction. The present paper is primarily our effort to answer question A, to provide a self-contained minimal formulation of QM from which the answers to questions B and C follow in the course of research, i.e. pursuing \lq what physicists do'. Question D was an attempt to list many of the items we have termed \lq excess baggage', namely many issues that do not need to be addressed in order to arrive at a satisfactory minimal formulation. It is not that we wish to censor what are legitimate subjects of study, but rather that in our view the first step in quantum foundations is defining the theory clearly and succinctly in order to be able to test, judge and critique it. Many treatments, however, replace or combine this goal with many others, for example the search for alternatives to Hilbert space QM, a move for which the physical motivation is often unclear. We have thus chosen to classify such searches under item D of our Introduction and suggested that in many cases they might represent excess baggage.

In conclusion, the authors would like to express the admittedly quixotic hope that the subject of Foundations of QM may attain the same status as its cousin, Foundations of CM.

 \vspace{10mm}
\numberwithin{equation}{section}
\appendix
\section{\label{A} Set Theory, Classical Logic and Probability Theory}

In this appendix we provide a brief summary of set theory, classical (Aristotelian) logic and classical probability theory, and we show how the three are formally related.

\vskip10pt

\noindent
\underline{Set Theory}

For simplicity we consider a discrete set $\Omega$ of $N$ elements $x\in \Omega$. The subsets $A,B,...$ of $\Omega$ form a set $\mathcal{B}(\Omega)$ 
of sets (in set theory, a \emph {field} of sets) 
for which the operations of union $\cup$, intersection $\cap$ and complement $\sim$ obey the axioms  of set theory:

\begin{subequations}
\label{sets}
\begin{eqnarray}
A\cup \oslash = A & A\cap \Omega = A, \\
A\cup \sim A = \Omega & A\cap \sim A = \oslash,\\
A\cup B = B\cup A  &  A\cap B = B\cap A, \\
A\cup (B\cup C) = (A\cup B)\cup C, & A\cap (B\cap C) = (A\cap B)\cap C,\\
A\cup (B\cap C) = (A\cup B)\cap (A\cup C), & \;\; A\cap (B\cup C) = (A\cap B)\cup (A\cap C),\label{setdist}
\end{eqnarray}
\end{subequations}	
where $\oslash$ is the empty subset of $\Omega$.


\noindent
\underline{Classical Logic}

The subsets $A,B,...$ can also be considered as logical propositions,  in which case the operations of set theory become logical operations

\begin{subequations}
\label{setlog}
\begin{eqnarray}
\cup\;\; \longrightarrow \;\; \vee\;\; \text{disjunction\;\;(or)},\label{q-1}  \\
\cap\;\; \longrightarrow \;\; \wedge\;\; \text{conjunction\;\;(and)},\label{q-2}  \\
\sim\;\; \longrightarrow \;\; \neg\;\; \text{negation\;\;(not)},\label{q-3}\\
\Omega\;\; \longrightarrow \;\; \text{T}\;\; (\text{true}),\label{q-4}\\
\oslash\;\; \longrightarrow \;\; \text{F}\;\; (\text{false}).\label{q-5}
\end{eqnarray}
\end{subequations}

\noindent
Under the replacements \eqref{setlog}, \eqref{sets} become the usual axioms of propositional calculus
\begin{subequations}
\label{clogic}
\begin{eqnarray}
A \vee \text{F} = A & A \wedge \text{T} = A, \\
A \vee\sim A = \text{T} & A \wedge\sim A = \text{F},\\
A \vee B = B \vee A  &  A \wedge B = B \wedge A, \\
A \vee (B \vee C) = (A \vee B) \vee C, &\;\; A \wedge (B \wedge C) = (A \wedge B) \wedge C,
\end{eqnarray}
\begin{eqnarray}\label{clogdist}
A \vee (B \wedge C) = (A \vee B) \wedge (A \vee C), & \;\; A \wedge (B \vee C) = (A \wedge B)\vee (A \wedge C).
\end{eqnarray}
\end{subequations}	
\noindent In particular \eqref{setdist} becomes the distributive law \eqref{clogdist}. The set ${\mathcal{B}(\Omega)}$ of $2^N$ propositions forms a Boolean algebra under the logical operations.
On this algebra we can define truth functions $\mathcal{T}(A)$ with values 1 (True) and 0 (False).
Such truth functions must agree with the standard truth tables for the logical functions,
which imply the algebraic relations

\begin{subequations}
\label{tfbool}
\begin{eqnarray}
\mathcal{T}(\neg A) = 1 - \mathcal{T}(A),  \\
\mathcal{T}(A\wedge B) = \mathcal{T}(A) \mathcal{T} (B),  \\
\mathcal{T}(A\vee B) = \mathcal{T}(A) + \mathcal{T}(B) - \mathcal{T}(A) \mathcal{T} (B), \label{dist2}
\end{eqnarray}
\end{subequations}

\noindent and must also satisfy  $\mathcal{T}(\Omega)=1$,  $\mathcal{T}(\oslash)=0$.
We shall refer to these equations as \lq truth table relations'.

Let us consider in particular those subsets of $\Omega$ containing only one member, that is sets of the form $\{x\}$ where $x\in\Omega$.  We may call them \emph{atomic} sets, and the corresponding logical propositions atomic propositions.  Then by applying \eqref{tfbool} we find that any truth function must assign the value 1 to some atomic proposition and 0 to all the others.  We shall denote the truth function that assigns 1 to a particular $\{x\}$ by ${\cal T}_x$.  Then for $x, y \in \Omega$

\begin{equation}
{\cal T}_x(\{y\}) = 1  \;\;\; \text{if} \;\;\;  y=x, \;\;\; \text{otherwise}\;\;\; 0.
\end{equation}

\noindent Again  applying \eqref{tfbool}, we see that for any $A\in \mathcal{B}(\Omega)$,

\begin{equation}\label{Txdef}
{\cal T}_x(A) = 1  \;\;\; \text{if} \;\;\;  x\in A, \;\;\; \text{otherwise}\;\;\; 0.
\end{equation}

\noindent We may say that $x$ is the \lq\lq source of truth" for the atomic truth function ${\cal T}_x$.

\noindent
\underline{Probability Theory}

Truth functions can be generalized by introducing a \emph{probability function}
$\mathcal{P}(A)$ from $\mathcal{B}(\Omega)$ to the unit interval $[0,1]$.  One first defines a \emph{measure} as a function from $\mathcal{B}(\Omega)$ to the interval $[0,\infty]$, which satisfies the linearity condition for countable sets of \emph{disjoint}
subsets,

\begin{equation} \label{kolminf}
\mathcal{P}(A^{(1)}\vee A^{(2)}\vee ... ) =  \mathcal{P}(A^{(1)})+ \mathcal{P}(A^{(2)})+..., \; \;\; \text{whenever}\;\; A^{(i)}\wedge A^{(j)} = \oslash \;\;\text{for}\;\;i\neq j.
\end{equation}

 \noindent A \emph{probability measure} or \emph{probability function} (classically, these two ideas can be identified) is a measure which satisfies the additional condition

\begin{equation} \label{probmeas}
\mathcal{P}(\Omega)=1.
\end{equation}

\noindent (The relation $\mathcal{P}(\oslash)=0$ is already implied by \eqref{kolminf}).
In the context of probability theory, the set $\Omega$ is the \lq sample space' of the probability measure and the elements $A^{(i)}\in \mathcal{B}(\Omega)$ are called \lq events'. It can be shown that for any two events $A,B$, \eqref{kolminf} implies the relation

\begin{equation} \label{kolm2}
\mathcal{P}(A \vee B) = \mathcal{P}(A) + \mathcal{P}(B) - \mathcal{P}(A \wedge B).
\end{equation}
\noindent The converse is true for a finite $\Omega$.  We shall sometimes refer to \eqref{kolminf} and \eqref{probmeas} as the \lq Kolmogorov conditions', and to \eqref{kolm2} as the \lq Kolmogorov overlap equation'.

One notices a similarity between \eqref{kolm2} and \eqref{dist2}. Indeed, in \cite[Appendix A]{fh1} the authors explained in what sense the probability function $\mathcal{P}$ can be thought of as a \lq distributed truth function'.

Let us mention one feature of probability functions which is often obscure, namely that (in a finite-dimensional sample space) defining a probability function that assigns a relative likelihood to the members of a sample space does not alter the fact that one and only one of its members is true, while all the others are false.

It should be noted, moreover, that our definitions of probability and truth are formal ones, and they are thus consistent with either a frequentist or a Bayesian approach to probabilities. At this stage we are not inquiring into the relationship of probabilities to the \lq real world', which is where such distinctions arise. The connection between truth and probability explored above exists already on the formal level and is therefore independent of any real-world interpretation of probability.

\vspace{10mm}

\section{\label{B} Noncontextual Network Theorem}

 Suppose we have a noncontextual network $\cal{N}$ of probability functions, i.e. a set of functions 
 $P_{\cal F}$ with sample spaces $\mathcal{F}$, such that if $A$ belongs both to ${\cal F}_1$ and to
${\cal F}_2$, then
\begin{equation}\label{probb0}
P_{{\cal F}_1}(A) = P_{{\cal F}_2}(A).
\end{equation}
\noindent We show here that the probability functions $P_{\cal F}$ must satisfy the Born Rule \eqref{Born2}, for some density matrix $\rho$.

To prove the theorem we first define a quantum probability measure ${\mathcal M}_{\cal N}$. For each $A \in Q(\mathcal{H})$, let ${\cal F}_A$ be the framework consisting of 1, $\oslash$, $A$, and $\neg A$.  Then we define
\begin{equation}\label{probb00}
{\cal M}_{\cal N}(A) = P_{{\cal F}_A}(A).
\end{equation}

\noindent Is ${\mathcal M}_{\cal N}$ a quantum probability measure? That depends on whether the additivity condition

\begin{equation} \label{Wmackinf}
{\mathcal M}_{\cal N}(A^{(1)}\vee A^{(2)}\vee ... ) = {\mathcal M}_{\cal N}(A^{(1)}) +  {\mathcal M}_{\cal N}(A^{(2)}) + ..., \text{when}\;\; A^{(i)} \perp A^{(j)}\;\; \text{for} \; i \neq j,
\end{equation}

\noindent is satisfied.  So let $\{A\} = \{A_1, A_2, ...\}$ be a finite or countably infinite set of properties that are mutually orthogonal, that is $A_i \perp A_j$ for $i\neq j$ as in the condition  \eqref{Wmackinf} ; and let $\textrm{Sp}(\{A\})$ be the span of
all the members of $\{A\}$.  If we define $A_0 = \neg \textrm{Sp}(\{A\})$, then $S_{\{A\}} = \{A_0, A_1, A_2,...\}$ is a sample space. This sample space is the basis of a framework which we shall call ${\cal F}_{\{A\}}$.
Since $P_{{\cal F}_{\{A\}}}$ is a probability function, \eqref{kolminf} above tells us that
\begin{equation}\label{probb1}
P_{{\cal F}_{\{A\}}}(A^{(1)}\vee A^{(2)}\vee ...) = P_{{\cal F}_{\{A\}}}(A_1) + P_{{\cal F}_{\{A\}}}(A_2) + ... \,.
\end{equation}

\noindent But since any $A\in{\cal F}_{\{A\}}$ belongs to both ${\cal F}_{\{A\}}$ and ${\cal F}_{A_i}$, we can write \eqref{probb1} as
\begin{equation}\label{probb2}
P_{{\cal F}_{\{A\}}}(A^{(1)}\vee A^{(2)}\vee ...) = P_{{\cal F}_{A_1}}(A_1) + P_{{\cal F}_{A_2}}(A_2) + ...,
\end{equation}

\noindent in view of \eqref{probb0}.  Then using \eqref{probb00}, we have
\begin{equation}
{\cal M}_{\cal N}(A^{(1)}\vee A^{(2)}\vee ...) = {\cal M}_{\cal N}(A_1) + {\cal M}_{\cal N}(A_2) + ...,
\end{equation}

\noindent which is \eqref{Wmackinf}.  Therefore ${\mathcal M}_{\cal N}$ is a quantum probability measure.  Its normalization ${\mathcal M}_{\cal N}({\cal H}) = 1 $ can be deduced from the completeness of the sample space in any framework.  Then from Gleason's theorem we get the Born rule \eqref{Born2}.

\vspace{10mm}

\section{\label{C} Kochen's proof of an important theorem}

In \cite[Sec. 8.1]{kochen2015}, a theorem is stated and proved:  that given a state $p$ and a property $y$ such that $p(y)\neq 0$, a unique new state $p_y$ exists such that $p_y(x) = p(x\wedge y)/p(y)$ for any $x$ belonging to some $\sigma$-algebra $B$ that contains $y$; and that the density operator of $p_y$ is given uniquely by $w_y = ywy/Tr(ywy)$ where $w$ is the density operator associated with $p$.  The proof is valid but extremely terse.  For better comprehension we give here an expanded version of the theorem and the proof.

\nopagebreak

\subsection{\label{C.1} Preliminary Definitions}

In \cite[Sec. 3.1]{kochen2015}  we read,

``A \emph{state} of a system with a $\sigma$-complex of properties $Q$ is a map $p: Q -> [0,1]$ such that the restriction of $p$ to any $\sigma$-algebra $B$ in $Q$ is a probability measure on $B$.''

Looking higher in the same section, we see that a probability measure on $B$ is a function $p: B -> [0,1]$  such that

\begin{equation}\label{pI}
p(I) = 1
\end{equation}

\noindent  where $I$ is the identity operator, and
\begin{equation}
p(a_1 \vee a_2 \vee...) = p(a_1) + p(a_2)  + ...
\end{equation}

\noindent for any mutually disjoint set of $a_i$ all belonging to $B$.

But if the $a_i$ are properties, disjoint simply means orthogonal; and for orthogonal properties $x_i \vee x_j = x_i + x_j$.  So the definition of a state can be rewritten

A \emph{state} of a system with a $\sigma$-complex of properties $Q$ is a map $p: Q -> [0,1]$ such that \eqref{pI} holds and
\begin{equation}\label{psum}
p(x_1 + x_2 + ...) = p(x_1) + p(x_2) +...
\end{equation}

\noindent for any mutually orthogonal $x_1, x_2, ...$.  (It is trivial that such a set generates a $\sigma$-algebra.)

Now looking lower, we see that if $Q = Q(\cal H)$ where $\cal H$ is a Hilbert space of dimension $>2$, Gleason's theorem tells us that for any state
$p$ there exists a unique density operator $w$ (nonnegative Hermitian operator of trace 1) on $\cal H$ such that
\begin{equation}\label{Bsimp}
p(x) = Tr(w x)
\end{equation}

\noindent for all $x\in Q$.

\subsection{\label{C.2} Statement of Theorem}

In \cite[Sec. 8.1]{kochen2015}, Kochen states and proves the following theorem.  (We replace some of his notation with our own.)

If $p$ is a state on $Q(\cal H)$ and $y\in Q(\cal H)$ such that $p(y) \neq 0$, then there exists a unique state $p_y$ conditioned on $y$.  If
$w$ is the density operator corresponding under \eqref{Bsimp} to $p$, then $ywy/Tr(ywy)$ corresponds to the state $p_y$.

To understand this statement one must look back a couple of paragraphs to the general $\sigma$-complex $Q$:

"Let $p$ be a state on a $\sigma$-complex Q and $y\in Q$ such that $p(y)\neq 0$.  By a \emph{state conditioned on $y$} we mean a state
$p_y$ such that for every $\sigma$-algebra $B$ in $Q$ containing $y$ and every $x\in B$,
\begin{equation}\label{wedge}
p_y(x) = p(x\wedge y)/p(y)."
\end{equation}

This permits us to \emph{restate the theorem} as follows:

Let it be given that $p$ is a state on $Q(\cal H)$ as defined in the previous section, and that $(p,w)$ satisfy \eqref{Bsimp}.   Let it also be given that $y\in Q(\cal H)$ and that
\begin{equation}\label{pneq0}
p(y) \neq 0.
\end{equation}

Then there exists a state $p_y$ on $Q(\cal H)$ such that (5) holds for every $x\in Q(\cal H)$ that belongs to some
$\sigma$-algebra $B$ that also contains $y$. Moreover, there exists only one such state, and it is given by the formula
\begin{equation}\label{x:y}
p_y(x) = Tr(w_y x)
\end{equation}

\noindent where
\begin{equation}\label{wy}
w_y = ywy/Tr(ywy).
\end{equation}

[It is important to note that the assertion that $p_y$ is a state implies that $p_y(x)$ is defined for \emph{every} property $x$, whether or not it commutes with $y$.  But \eqref{wedge} is asserted only for certain properties $x$.  In fact, one might as well say that \eqref{wedge} is asserted for all $x$ compatible with $y$, since if $x$ and $y$ commute one can easily construct a $\sigma$-algebra containing them both.

It is also essential that \eqref{pneq0} be assumed, since otherwise the right side of \eqref{wedge} could be 0/0.

Finally, \eqref{x:y} is meant to hold for all $x$, not only those satisfying \eqref{wedge}.]

\subsection{\label{C.3} Proof of Theorem (existence)}

The proof of the theorem is in two halves.  First, we must show that there exists a state $p_y$ such that \eqref{wedge} holds for all $x$ that commutes with $y$.  And second, we must show that any such $p_y$ must satisfy \eqref{x:y} and \eqref{wy}.  In both halves it is given to begin with that $p$ is a state and that $y$ is a property satisfying \eqref{pneq0}.

To show that a state $p_y$ exists with the desired feature, we simply exhibit one as defined by \eqref{x:y} and \eqref{wy}.  We must then show that $p_y$ so defined is a state, and also that it satisfies \eqref{wedge} for all $x$ compatible with $y$.  The latter statement is easily proved.  Combining \eqref{x:y} and \eqref{wy}, we have
\begin{equation}\label{pyx}
p_y(x) = Tr(ywyx)/Tr(ywy)
\end{equation}

\noindent for all $x$.  The denominator is equal to $Tr(wyy) = Tr(wy) = p(y)$, which is nonzero by \eqref{pneq0}; therefore \eqref{pyx} is defined.  Then if we specialize to $x$ commuting with $y$, the numerator in \eqref{pyx} can be written as $Tr(ywxy) = Tr(wxyy) = Tr (wxy) = p(xy)$; but for commuting $x$ and $y$, $xy$
is the same as $x\wedge y$ and so \eqref{wedge} is satisfied.

It remains to show that $p_y$ as defined by \eqref{pyx} is a state.  This means verifying \eqref{pI} and \eqref{psum} with $p_y$ in place of $p$. As for \eqref{pI}, we have from \eqref{pyx}
\begin{equation}
p_y(I) = Tr(ywyI)/Tr(ywy) = Tr(ywy)/Tr(ywy) = p(y)/p(y) = 1,
\end{equation}

\noindent again appealing to \eqref{pneq0}. As for \eqref{psum}, the equation
\begin{equation}
p_y(x_1 + x_2 +...) = p_y(x_1) + p_y(x_2)  +...
\end{equation}

\noindent follows directly from \eqref{x:y} since the right side of \eqref{x:y} is linear in $x$. Therefore $p_y$ is a state.

\subsection{\label{C.4} Proof of Theorem (uniqueness)}

In this section we assume as before that $p$ is a state and that $y$ is a property obeying \eqref{pneq0}.  We assume also that $p_y$ is some state for which
\eqref{wedge} holds whenever $x$ commutes with $y$.  Then from Gleason's theorem we know that a density operator $w_y$ exists satisfying \eqref{x:y} for all $x$.

But we do not assume \eqref{wy}; we regard the form of $w_y$ as unknown.  We shall then derive an expression for $p_y(x)$, holding for any 1-dimensional $x$, from which $w_y$ has been eliminated.  (This is the heart of Kochen's proof. It shows that the state $p_y$ is unique, since its
values on arbitrary $x$ can be obtained from those on 1-dimensional $x$ by linearity.)

If $x$ is 1-dimensional, it can be written as $| \phi\!><\!\phi |$ for some ket $|\phi\!>$ for which $<\!\phi | \phi\!> = 1$.  Then \eqref{x:y} becomes
\begin{equation}\label{Bn}
p_y(x) = <\!\phi | w_y |\phi\!>.
\end{equation}

Since $y$ and its complement
$y^{\perp}$ span $\cal H$,  we have
\begin{equation}\label{phph}
|\phi\!> = y |\phi\!> + \; y^{\perp}|\phi\!>.
\end{equation}

\noindent By substituting \eqref{phph} for  $| \phi\!>$ in \eqref{Bn}, we obtain $p_y$ as the sum of four terms, three of which contain the factor
$w_y y^{\perp}| \phi\!>$ or else $<\!\phi | y^{\perp} w_y$.   These three vanish, for the following reason.  Since $y$ and $y^{\perp}$ commute, we may substitute $y^{\perp}$ for $x$ in \eqref{wedge}, obtaining
\begin{equation}\label{yperp}
p_y(y^{\perp}) = p( y^{\perp}\wedge y)/p(y) = p(\emptyset)/p(y) = 0,
\end{equation}

\noindent where $\emptyset$ is the empty set.  ($p(\emptyset)$ must vanish, otherwise \eqref{psum} would lead to a contradiction.)  On the other hand, since \eqref{x:y} applies to all $x$ we can apply it to $y^{\perp}$, obtaining
\begin{equation}\label{Bperp}
p_y(y^{\perp}) = Tr(w_y y^{\perp}),
\end{equation}

\noindent and comparing \eqref{Bperp} to \eqref{yperp} we have
\begin{equation}\label{Trwy0}
Tr(w_y y^{\perp}) = 0.
\end{equation}

Now from \eqref{Trwy0} it follows that $w_y y^{\perp} |\phi\!> = 0$.  This implication is not trivial, and we shall give the reasoning as a lemma.

Lemma 1: If $v$ is a density operator and $z$ is a projector, then either $vz = 0$ or $Tr(vz) > 0$.

Proof:  $z$ can be decomposed into a sum of one-dimensional projectors $z_k$, and the lemma will be true of $z$ if it is true of each $z_k$.
Therefore without loss of generality we can take $z$ to be one-dimensional: $z = |\xi\!\!><\!\!\xi|$ where $<\!\!\xi|\xi\!\!> = 1.$

Express $v$ in matrix form so that it is diagonal: $v_{ab} = n_a \delta_{a,b}$ where all $n_a \geq 0$ and $\Sigma n_a = 1$.
Then $|\xi\!\!>$ is a column matrix whose $a$'th element is $\xi_a$, and $\Sigma |\xi_a|^2 = 1.$

We then have $z_{ab} = \xi_a \xi_b^*$, and $(vz)_{ab} = n_a \xi_a \xi_b^*.$  Hence $Tr((vz) = \Sigma n_a \xi_a \xi_a^*.$   Each term of the trace is nonnegative, and therefore $Tr(vz) > 0$ unless, for each $a$, $n_a \xi_a \xi_a^* = 0$.  But in that case, $|n_a \xi_a|^2 = 0$, hence $n_a \xi_a = 0$ and therefore $n_a \xi_a \xi_b^* = 0$ for all $b$; therefore $vz = 0$.  The lemma is proved.

Now in Lemma 1 take $v = w_y$, $z = y^{\perp}$, and we see that \eqref{Trwy0} implies $w_y y^{\perp} = 0$.  Therefore in the expansion of \eqref{Bn} the three terms containing $w_y y^{\perp} |\phi\!>$ or its dual vanish, and we are left with
\begin{equation}\label{pwy}
p_y(x) = <\!\phi |y w_y y|\phi\!>.
\end{equation}

To eliminate $w_y$, we shall first prove a second lemma:

Lemma 2:
\begin{equation}\label{L2}
<\!\phi |y w_y y|\phi\!> = <\!\phi |y|\phi\!> Tr(w_y u),
\end{equation}

\noindent where $u$ is the projector onto $y|\phi\!>$.  (Thus
\begin{equation}
u = \nu y|\phi\!><\!\phi |y\label{unu}
\end{equation}

\noindent for some nonzero real number $\nu$.)

Proof:

If $y|\phi\!>=0$ then $u=0$ and both sides of \eqref{L2} vanish.

Otherwise $u\neq 0$ and let $\mu = <\!\phi |y|\phi\!> = <\!\phi |yy|\phi\!>$ .  We have
\begin{equation}
u = uu = \nu^2 y|\phi\!><\!\phi |y\, y|\phi\!><\!\phi |y = \nu^2 \mu\, y|\phi\!><\!\phi |y = \nu\mu u
\end{equation}

\noindent so that $\nu\mu = 1$ and \eqref{unu} becomes
\begin{equation}
u = (1/ \mu) y|\phi\!><\!\phi |y.
\end{equation}

Let $|\chi_0\!> = y|\phi\!>/\sqrt{\mu}$ so that $<\!\chi_0 |\chi_0\!> = 1.$  Then
\begin{equation}\label{uch}
u = |\chi_0\!><\!\chi_0 |.
\end{equation}

Let $|\chi_0\!>, |\chi_1\!>, |\chi_2\!>, ...$ be a complete orthonormal basis for $\cal H$.  Then
\begin{equation}
Tr(w_y u) = Tr(w_y uu) = Tr(uw_yu) = \Sigma_i <\!\chi_i | u w_y u |\chi_i \!> .
\end{equation}

\noindent But by \eqref{uch} we have
\begin{equation}\label{chdel}
<\!\chi_i | u w_y u |\chi_i \!> = <\!\chi_i | \chi_0\!><\!\chi_0 |w_y|\chi_0\!><\!\chi_0 |\chi_i\!> = \delta_{i,0}<\!\chi_0 |w_y|\chi_0\!>
\end{equation}

\noindent and therefore
\begin{equation}\label{wch}
Tr(w_y u) = <\!\chi_0| w_y | \chi_0\!> =  <\!\phi |y w_y y| \phi\!> /\mu
\end{equation}

\noindent which is just \eqref{L2}.  Lemma 2 is proved.

Therefore \eqref{pwy} becomes
\begin{equation}\label{pyxw}
p_y(x) = <\!\phi |y|\phi\!> Tr(w_y u).
\end{equation}

Now, comparing the projectors $u$ and $y$, we have
\begin{equation}
y| \chi_0\!> = y(y|\phi\!>\!/\sqrt{\mu}) = y|\phi\!>\!/\sqrt{\mu} = | \chi_0\!> = u | \chi_0\!>
\end{equation}

\noindent whereas $u | \chi_i\!> = 0$ for any $i\neq 0$.  Therefore $<\!\chi_i |y - u |\chi_i \!\!>\; \geq\! 0$ for all $i$, or simply $y\geq u$.
Hence $u$ and $y$ commute and
\begin{equation}
u \wedge y = u.
\end{equation}

\noindent It follows that (taking $B$ as the $\sigma$-algebra generated by $u$ and $y$) $u$ may be substituted for $x$ in \eqref{wedge}, yielding

\begin{equation}\label{ppp}
p_y(u) = p(u\wedge y) / p(y) = p(u) / p(y)
\end{equation}

\noindent which exists by \eqref{pneq0}.  Likewise, \eqref{x:y} - postulated to hold for \emph{all} $x$ - holds for $u$ in place of $x$:
\begin{equation}
p_y(u) = Tr(w_y u)\label{pwu}
\end{equation}.

\noindent Comparing \eqref{pwu} with \eqref{ppp}, we obtain
\begin{equation}
Tr(w_y u)  =  p(u) / p(y)
\end{equation}

\noindent so that \eqref{pyxw} becomes
\begin{equation}\label{puy}
p_y(x) =  <\!\phi |y|\phi\!>  p(u)/p(y).
\end{equation}

This is the promised expression for $p_y(x)$ that does not involve $w_y$.  (We have nowhere invoked \eqref{wy}.)  However, it holds only for 1-dimensional $x$ (of the form $| \phi\!><\!\phi |$).  To show that \eqref{puy} suffices to define the whole state $p_y$, we observe that \emph{any} $x$ is the sum of 1-dimensional projectors and that \eqref{x:y} is linear in $x$ \emph{no matter what form} $w_y$ takes.  Hence \eqref{puy} establishes the uniqueness of $p_y$ satisfying the conditions at the start of this section.

\subsection{\label{C.5}Unique form of density matrix}

In Section 3 we showed that a legitimate state $p_y$ exists, satisfying the conditions imposed for conditioning $p$ on $y$,  whose density operator is given by \eqref{wy}.  In Section 4 we showed that no other state can satisfy these conditions.  But Gleason's theorem tells us that a given state can have only one density operator.  Hence in fact \eqref{wy} does describe the density operator associated with the state $p_y$ whose uniqueness is shown in Section 4.

This completes the entire proof of Kochen's theorem in \cite[Sec. 8.1]{kochen2015}.



\end{document}